\documentclass[preprint]{elsarticle}

\usepackage{graphicx,rotating}
\usepackage{epsfig}
\usepackage{amssymb}

\journal{Nuclear Physics A}

\newcommand{\ba}{\begin{eqnarray}}
\newcommand{\ea}{\end{eqnarray}}

\begin{document}

\begin{frontmatter}

\title{Cluster structure of $^{20}$Ne: Evidence for ${\cal D}_{3h}$ symmetry}
\author{R. Bijker}
\address{Instituto de Ciencias Nucleares, 
Universidad Nacional Aut\'onoma de M\'exico, \\
Apartado Postal 70-543, 04510 Cd. de M\'exico, M\'exico}
\ead{bijker@nucleares.unam.mx}

\author{F. Iachello}
\address{Center for Theoretical Physics, Sloane Laboratory, Yale University, \\ 
New Haven, CT 06520-8120, U.S.A.}
\ead{francesco.iachello@yale.edu}

\begin{abstract}
We study the cluster structure of $^{20}$Ne and show that the available
experimental data can be well described by a bi-pyramidal structure with ${\cal D}_{3h}$ symmetry. 
Strong evidence for the occurrence of this symmetry comes
from the observation of all nine expected vibrational modes (3 singly
degenerate and 3 doubly degenerate) and of six (singly degenerate) double
vibrational modes. $^{20}$Ne appears to be another example of the simplicity
in complexity program, in which simple spectroscopic features arise out of a
complex many-body system.
\end{abstract}


\end{frontmatter}

\section{Introduction}

The cluster structure of light nuclei has a long history dating back to
Wheeler \cite{wheeler} and Hafstad and Teller \cite{hafstad} who noted that
the binding energies of $N=Z$ nuclei from $^{8}$Be to $^{32}$S lie on a
straight line when plotted against the number of $\alpha -\alpha $ adjacent
bonds. This work was followed by Dennison \cite{dennison} and Kameny \cite%
{kameny}. In 1965, Brink \cite{brink} suggested specific cluster
configurations for nuclei composed of $k$ $\alpha $-particles, henceforth
referred as $k\alpha $ nuclei. Ground state configurations from $^{12}$C to $%
^{28}$Si were studied \cite{brink2,weiguny1,weiguny2}.
Additional configurations for nuclei in the s-d shell were also investigated 
\cite{hauge}. In a parallel development, the connection between the shell
model and the cluster model was investigated by Wildermuth and Kannelopoulos 
\cite{wildermuth}, as well as by the Japanese school 
\cite{ikeda,fujiwara,horiuchi1,horiuchi2}. A recent review is given in 
\cite{vonOertzen-review}.

Most studies of cluster structures in light nuclei have been confined to $%
^{8}$Be, $^{12}$C and $^{16}$O, especially $^{12}$C for which measurements
of new rotational states have been recently done \cite{freer1,freer2,kirsebom,marin}. Evidence has been presented for a triangular configuration in $^{12}$C ($k=3$) with symmetry ${\cal D}_{3h}$ \cite{bijker1,bijker2}, and a tetrahedral configuration in $^{16}$O ($k=4$) with symmetry $T_{d}$ \cite{robson,bijker3,bijker4}. An
important question is the extent to which cluster structures appear in
heavier systems, in particular in nuclei with $16<A<40$, and, if so, what is
their structure and symmetry. Studies within the framework of the
Brink-Bloch model were done in the early 1970's \cite{brink2,weiguny1,weiguny2}, and some evidence was found on the basis of a
limited amount of data. Since then, a large number of experimental data has
been accumulated and it appears therefore of interest to revisit the
question of clustering in $16<A<40$ nuclei in light of the new experimental
information.

Nuclei with $16<A<40$ have also been the subject of extensive studies within
the framework of the spherical shell-model \cite{deshalit}, of the $SU(3)$
Elliott model \cite{elliott} and its generalization to symplectic $Sp(6,R)$ 
\cite{rosensteel}, \cite{rowe1}, and of the collective quadrupole model \cite%
{bohr} and its generalization to odd nuclei \cite{nilsson}. In very recent
years also large scale shell-model calculations have become available \cite%
{caurier}.

In this article, we study the cluster structure of $^{20}$Ne ($k=5$).
Several configurations are possible for $k=5$ particles. Two of these are
the bi-pyramidal configuration suggested in \cite{brink2,weiguny1,weiguny2} with symmetry ${\cal D}_{3h}$ (or ${\cal D}_{3}$) and the distorted 
body-centered tetrahedral configuration suggested in \cite{hauge} with
symmetry ${\cal D}_{2d}$. In view of the fact that all microscopic calculations
within the framework of the Brink-Bloch model, of the shell model and, very
recently, of Density Functional Theory (DFT) \cite{vretenar} suggest a
bi-pyramidal structure, we concentrate in this article on this structure. We
first develop in Sect. 2 all formulas needed to describe ground state
properties. These formulas include moment of inertia, r.m.s. radius, form
factors in electron scattering and electromagnetic transition rates. In
Sect. 3 we compare the results so obtained with experiment and in Sect. 4
with other models. In Sect. 5, we consider the vibrations of the
bi-pyramidal structure and show that all nine of them have been observed. In
the study of the vibrations the full power of group theoretical methods
comes into play, as we identify the vibrations by the expected irreducible
representations of ${\cal D}_{3h}$. The observation of all vibrations provides
strong evidence for the occurrence of ${\cal D}_{3h}$ symmetry in $^{20}$Ne. In
Sect. 6 the electromagnetic properties of the vibrations are investigated.
In Sect. 7 we discuss the double vibrational spectrum and show that six of
these double vibrational bands have been observed, thus providing further
evidence for the occurrence of ${\cal D}_{3h}$ symmetry in $^{20}$Ne. Finally,
Sects. 8-10 contain a brief comparison with other descriptions and Sect. 11
the conclusions.

\section{Bi-pyramidal cluster configuration}

\subsection{Geometric structure}

\begin{figure}
\centering
\vspace{15pt}
\setlength{\unitlength}{1pt}
\begin{picture}(250,200)(0,20)
\thicklines
\put( 70, 90) {\circle*{8}} 
\put( 30,110) {\circle*{8}}
\put(110,110) {\circle*{8}}
\put( 80,185) {\circle*{8}}
\put( 80, 50) {\circle*{8}}
\put( 20,115) {$3$}
\put(115,115) {$2$}
\put( 70,185) {$4$}
\put( 65, 45) {$5$}
\put( 75, 82) {$1$} 
\put( 30,110) {\line( 2, 3){50}}
\put( 30,110) {\line( 5,-6){50}}
\put( 30,110) {\line( 2,-1){40}}
\put(110,110) {\line(-2, 5){30}}
\put(110,110) {\line(-2, -1){40}}
\put(110,110) {\line(-1, -2){30}}
\put( 80, 50) {\line(-1, 4){10}}
\multiput( 30,110)(5,0){16}{\line(1,0){2}}
\thinlines
\put( 80,105) {\vector(-2,-3){ 9}}
\put( 70, 90) {\vector( 2, 3){ 9}}
\put( 60,100) {$\beta_1$}
\put( 80,105) {\vector(-2,-3){ 35}}
\put( 80,105) {\vector( 0, 1){110}}
\put( 80,105) {\vector( 1, 0){ 70}}
\put( 78,220) {$\hat{z}$}
\put(155,102) {$\hat{y}$}
\put( 35, 45) {$\hat{x}$}
\multiput( 80,105)( 0,-4){4}{\circle*{1}}
\multiput( 80, 50)( 0, 4){8}{\circle*{1}}
\multiput( 80, 50)( 4, 0){15}{\circle*{1}}
\multiput( 80,185)( 4, 0){15}{\circle*{1}}
\put(133, 50) {\vector(0, 1){55}}
\put(133,105) {\vector(0,-1){55}}
\put(133,105) {\vector(0, 1){80}}
\put(133,185) {\vector(0,-1){80}}
\put(138,140) {$\beta_2$}
\put(138, 75) {$\beta_3$}
\put(190,170) {$\vec{r}_1=(\beta_1,\frac{\pi}{2},0)$}
\put(190,150) {$\vec{r}_2=(\beta_1,\frac{\pi}{2},\frac{2\pi}{3})$}
\put(190,130) {$\vec{r}_3=(\beta_1,\frac{\pi}{2},\frac{4\pi}{3})$}
\put(190,110) {$\vec{r}_4=(\beta_2,0,-)$}
\put(190, 90) {$\vec{r}_5=(\beta_3,\pi,-)$}
\end{picture}
\caption{A bi-pyramidal structure with ${\cal D}_3$ symmetry. For $\beta_2=\beta_3$ the structure has reflection symmetry with respect to the horizontal $xy$-plane, and the symmetry becomes ${\cal D}_{3h}$.}
\label{bipyramid}
\end{figure}
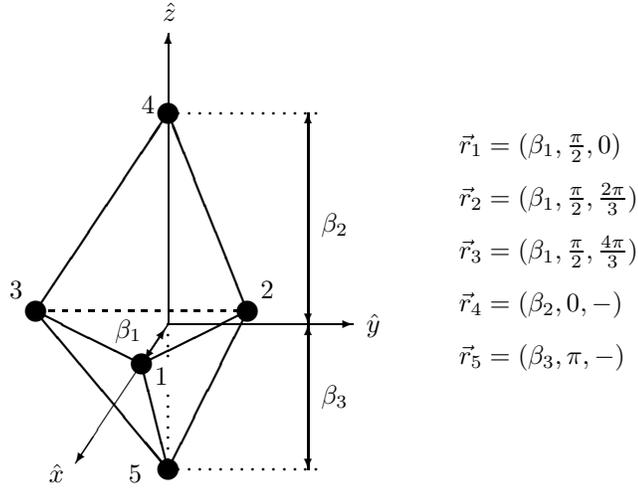

The most general geometric structure of a bi-pyramidal configuration with $%
D_{3}$ symmetry is given in terms of three length scales $\beta _{1},\beta
_{2},\beta _{3}$ (see Fig.~\ref{bipyramid}). The coordinates $(r_{i},\theta _{i},\phi _{i})$ of
the five constituent particles are%
\begin{eqnarray}
\left( r_{1},\theta _{1},\phi _{1}\right) &=& \left( \beta _{1},\frac{\pi }{2}%
,0\right) ~, 
\nonumber\\
\left( r_{2},\theta _{2},\phi _{2}\right) &=&\left( \beta _{1},\frac{\pi }{2}%
,\frac{2\pi }{3}\right) ~, 
\nonumber\\
\left( r_{3},\theta _{3},\phi _{3}\right) &=&\left( \beta _{1},\frac{\pi }{2}%
,\frac{4\pi}{3}\right) ~, 
\nonumber\\
\left( r_{4},\theta _{4},\phi _{4}\right) &=&\left( \beta _{2},0,-\right) ~, 
\nonumber \\
\left( r_{5},\theta _{5},\phi _{5}\right) &=&\left( \beta _{3},\pi,-\right) ~.
\end{eqnarray}%
For point particles the density normalized to 1 is%
\begin{equation}
\rho (\vec{r})=\frac{1}{5}\sum_{i=1}^{5}\delta \left( \vec{r}-\vec{r}%
_{i}\right) ~.
\end{equation}%
If the five particles are extended particles with gaussian distribution 
\begin{equation}
\rho _{\alpha }(\vec{r})=\left( \frac{\alpha }{\pi }\right) ^{3/2}e^{-\alpha
r^{2}} ~,
\end{equation}%
the density becomes%
\begin{eqnarray}
\rho (\vec{r}) &=&\frac{1}{5}\left( \frac{\alpha _{1}}{\pi }\right)
^{3/2}\sum_{i=1}^{3}\exp [-\alpha _{1}\left( \vec{r}-\vec{r}_{i}\right) ^{2}]
\nonumber \\
&&+\frac{1}{5}\left( \frac{\alpha _{2}}{\pi }\right) ^{3/2}\exp \left[
-\alpha _{2}(\vec{r}-\vec{r}_{4})^{2}\right]  \nonumber \\
&&+\frac{1}{5}\left( \frac{\alpha _{3}}{\pi }\right) ^{3/2}\exp \left[
-\alpha _{3}\left( \vec{r}-\vec{r}_{5}\right) ^{2}\right] ~,
\end{eqnarray}%
with three size scales $\alpha_{1}$, $\alpha_{2}$ and $\alpha_{3}$. This most
general bi-pyramidal configuration was considered in \cite{weiguny2} within
the framework of the $\alpha $-clustering model of Brink \cite{brink}. We
consider here the case of ${\cal D}_{3h}$ symmetry in which $\beta _{2}=\beta _{3}$%
, $\alpha _{2}=\alpha _{3}$. This configuration is considered in most
microscopic studies of the $\alpha $-clustering model of Brink \cite%
{weiguny1},\cite{weiguny2}. Generalization of the formulas given in the
following sections to the most general case is straightforward. The charge
distribution is obtained by multiplying $\rho \left( \vec{r}\right) $ by $Ze$
and the mass distribution by $Am$ where $Ze$ and $Am$ are the total charge and
mass. In the $\alpha $-clustering model $Ze/5=2e$ and $Am/5=4m$.

\subsection{Ground state properties: Formulas for ${\cal D}_{3h}$ symmetry}

From the density one can calculate all properties of a bi-pyramidal
configuration with ${\cal D}_{3h}$ symmetry, either directly, or simply noting that
the bi-pyramidal configuration with ${\cal D}_{3h}$ symmetry is composed of an
equilateral triangle in the $xy$-plane and a dumbbell in the $z$ direction.
Formulas for both configurations have been previously developed and are
reported in \cite{bijker-review}.

\subsubsection{Moments of inertia}

The moments of inertia are given by
\begin{eqnarray}
I_{z} &=& Am \int r^{2}drd\Omega \, \rho(\vec{r}) (x^{2}+y^{2}) ~,  
\nonumber\\
I_{x} &=& Am \int r^{2}drd\Omega \, \rho(\vec{r}) (y^{2}+z^{2}) ~,
\nonumber\\
I_{y} &=& Am \int r^{2}drd\Omega \, \rho(\vec{r}) (z^{2}+x^{2}) ~.
\end{eqnarray}
One obtains 
\begin{eqnarray}
I_{z} &=& \frac{3}{5}I_{z}(\Delta)
+\frac{2}{5}I_{z}(-) ~, 
\nonumber\\
I_{x} \;=\, I_{y} &=& \frac{3}{5}I_{x}(\Delta) 
+\frac{2}{5}I_{x}(-)  
\nonumber\\
&=& \frac{3}{5}I_{y}(\Delta) 
+\frac{2}{5}I_{y}(-) ~.
\end{eqnarray}
where $(\Delta)$ denotes an equilateral triangular configuration with ${\cal D}_{3h}$ symmetry \cite{bijker1,bijker2} and $(-)$ a dumbbell
configuration with $Z_{2}$ symmetry \cite{dellarocca}, with explicit
expressions 
\begin{eqnarray}
\frac{I_{z}}{m} &=& 12\left( \beta _{1}^{2}+\frac{1}{\alpha _{1}}\right) 
+ \frac{8}{\alpha_{2}} ~,
\nonumber\\
\frac{I_{x}}{m} \;=\; \frac{I_{y}}{m} &=& 6\left( \beta_{1}^{2}+\frac{2}{\alpha
_{1}}\right) +8\left( \beta _{2}^{2}+\frac{1}{\alpha _{2}}\right) ~.
\label{inertia}
\end{eqnarray}
From these one can calculate the inertia parameters
\begin{equation}
B \;=\; \frac{\hbar ^{2}/m}{2I/m} \;=\; \frac{20.7}{I/m} \mbox{ MeV} ~,
\label{coefb}
\end{equation}%
where $I/m$ is in fm$^{2}$ with $I$ the moment of inertia and $m$ the
nucleon mass. We note that since the bi-pyramid is a symmetric top, its
rotational energy levels are given by 
\begin{equation}
E_{rot}(L) \;=\; E_{0}+B_{x}L\left( L+1\right) +[B_{z}-B_{x}]K^{2} ~,
\end{equation}
where $L$ is the angular momentum and $K$ its projection on the intrinsic $z$%
-axis.

\subsubsection{Form factors}

Transition probabilities, charge radii and other electromagnetic properties of interest can be obtained from the transition form factors. For electric transitions the form factors are the matrix elements of the Fourier transform of the charge distribution
\ba
F_{\rm ch}(q) &=& \frac{1}{Ze} \int r^{2}drd\Omega \, \rho_{\rm ch}(\vec{r}) \, 
\mbox{e}^{i\vec{q} \cdot \vec{r}}  
\nonumber\\
&=& \frac{3}{5}F_{\rm ch}(q;\Delta)
+ \frac{2}{5}F_{\rm ch}(q;-) ~,
\nonumber\\
\rho_{\rm ch}(\vec{r}) &=& \frac{Ze}{5} \left( \frac{\alpha_{1}}{\pi} \right)^{3/2} \sum_{i=1}^{3} \mbox{e}^{-\alpha_{1}(\vec{r}-\vec{r}_{i})^{2}} 
\nonumber\\ 
&& + \frac{Ze}{5} \left( \frac{\alpha_{2}}{\pi} \right)^{3/2} \sum_{i=4}^{5} \mbox{e}^{-\alpha_{2}(\vec{r}-\vec{r}_{i})^{2}} ~.
\ea
The transition form factors from the ground state to states of the ground 
state rotational band labeled by angular momentum and parity $L^P$ and projection $K$ on the intrinsic symmetry-axis are given by \cite{hauge} 
\ba
F(i \rightarrow f;q) &=& \left< f \left| F_{\rm ch}(q) \right| i \right> 
\nonumber\\
&=& \frac{1}{5} \sqrt{4\pi} \sum_{i=1}^5  
j_L(qr_i) \, \mbox{e}^{-q^2/4\alpha_i} \, Y_{LK}(\theta_i,\phi_i) 
\nonumber\\
&=& \frac{3}{5} \sqrt{4\pi} \, j_L(q\beta_1) \, 
\mbox{e}^{-q^2/4\alpha_1} \, Y_{LK}(\frac{\pi}{2},0) \, \delta_{K,3\kappa} 
\nonumber\\
&& + \frac{2}{5} \sqrt{4\pi} \, j_L(q\beta_2) \, \mbox{e}^{-q^2/4\alpha_2} \,  \sqrt{\frac{2L+1}{4\pi}} \left( \frac{1+(-1)^L}{2} \right) \, \delta_{K,0} ~. 
\ea
Explicit expressions for transitions from the ground state to states of the ground state rotational band are obtained as
\begin{eqnarray}
F(0^+ \rightarrow 0^+;q) &=& 
\frac{3}{5} \, j_{0}(q\beta_{1}) \, \mbox{e}^{-q^{2}/4\alpha_{1}} 
+ \frac{2}{5} \, j_{0}(q\beta_{2}) \, \mbox{e}^{-q^{2}/4\alpha_{2}} ~, 
\nonumber\\
F(0^+ \rightarrow 2^+;q) &=& \frac{3}{5}\sqrt{\frac{5}{4}} \, j_{2}(q\beta_{1}) \, 
\mbox{e}^{-q^{2}/4\alpha_{1}} - \frac{2}{5}\sqrt{5} \, j_{2}(q\beta_{2}) \,  \mbox{e}^{-q^{2}/4\alpha_{2}} ~, 
\nonumber\\
F(0^+ \rightarrow 4^+;q) &=& \frac{3}{5} \frac{9}{8} \, j_{4}(q\beta_{1}) \, 
\mbox{e}^{-q^{2}/4\alpha_{1}} + \frac{2}{5} 3 \, j_{4}(q\beta_{2}) \, 
\mbox{e}^{-q^{2}/4\alpha_{2}} ~,
\nonumber\\
F(0^+ \rightarrow 3^-;q) &=& i\frac{3}{5}\sqrt{\frac{35}{8}} \, j_{3}(q\beta_{1}) \,
\mbox{e}^{-q^{2}/4\alpha_{1}} ~. 
\label{formfactors}
\end{eqnarray}
For the latter transition to the $3^-$ state only the first three $\alpha$-particles that constitute the central triangle contribute.

\subsubsection{Electromagnetic transitions and moments}

The transition probability $B(EL;0^+ \rightarrow L^P)$ can be extracted from the 
long wavelength limit of the transition form factor as
\ba
B(EL;0^+ \rightarrow L^P) \;=\; (Ze)^2 \, \frac{[(2L+1)!!]^{2}}{4\pi} \, \lim_{q\rightarrow 0} 
\frac{\left| F(0^+ \rightarrow L^P;q) \right|^{2}}{q^{2L}} ~,
\ea
to give
\ba
B(E2;0^+,0 \rightarrow 2^+,0) &=& \left( \frac{Ze}{5} \right)^2 \frac{5}{4\pi} 
\left( \frac{3}{2} \beta_1^2 - 2\beta_2^2 \right)^2 ~,
\nonumber\\
B(E4;0^+,0 \rightarrow 4^+,0) &=& \left( \frac{Ze}{5} \right)^2 \frac{9}{4\pi} 
\left( \frac{9}{8} \beta_1^4 + 2\beta_2^4 \right)^2 ~,
\nonumber\\
B(E3;0^+,0 \rightarrow 3^-,3) &=& \left( \frac{Ze}{5} \right)^2 \frac{7}{4\pi} 
\frac{45}{8} \beta_1^6 ~.
\ea
These expressions agree with the classical result for electric multipole radiation for a bi-pyramidal configuration of five $\alpha$-particles 
\ba
B(EL) \;=\; \sum_{M=-L}^{L} Q_{LM}^{\ast} Q_{LM} ~,
\ea
where $Q_{LM}$ are the multipole moments of the charge distribution
\ba
Q_{LM} &=& \int r^{L+2}Y_{LM}(\theta,\phi) \rho_{\rm ch}(\vec{r}) \, drd\Omega 
\nonumber\\
&=& \frac{3}{5} Q_{LM}(\Delta) 
+ \frac{2}{5}Q_{LM}(-) ~.
\ea

\paragraph{Quadrupole moment}

The intrinsic quadrupole moment is defined in terms of the multipole moment 
$Q_{20}$ as  
\ba
Q_0 \;=\; \sqrt{\frac{16\pi}{5}} \, Q_{20} \;=\; 
\left( -6\beta_{1}^{2}+8\beta_{2}^{2} \right) e ~.
\ea
The minus sign in front of $\beta_{1}^{2}$ reflects the fact that the triangular configuration is oblate, while the positive sign in front of $\beta_{2}^{2}$ reflects the fact that the dumbbell configuration is prolate. The other multipole moments vanish for the bi-pyramidal configuration, $Q_{2,\pm 1}=Q_{2,\pm 2}=0$. 
The $B(E2)$ value and the spectroscopic quadrupole moment of the $2^{+}$ state can be expressed in terms of the intrinsic quadrupole moment as
\begin{eqnarray}
B(E2;0^{+} \rightarrow 2^{+}) &=& \frac{5}{16\pi} \, Q_{0}^{2} ~, 
\nonumber\\
Q_{2^{+}} &=& -\frac{2}{7} \, Q_{0} ~.
\end{eqnarray}

\paragraph{Hexadecapole moment}

The intrinsic hexadecapole moment is defined in terms of the multipole moment $Q_{40}$ as \cite{horikawa}
\ba
H_0 \;=\; \sqrt{\frac{4\pi}{9}} \, Q_{40} \;=\; 
\left( \frac{9}{4}\beta_{1}^{4}+4\beta_{2}^{4} \right) e ~.
\ea
The other multipole moments vanish for the bi-pyramidal configuration, 
$Q_{4,\pm 1}=Q_{4,\pm 2}=Q_{4,\pm 3}=Q_{4,\pm 4}=0$. The $B(E4)$ value and the spectroscopic hexadecapole moment of the $4^{+}$ state are given by \cite{horikawa} 
\begin{eqnarray}
B(E4;0^{+} \rightarrow 4^{+}) &=& \frac{9}{4\pi} \, H_{0}^{2} ~,
\nonumber\\
H_{4^{+}} &=& H_{0} ~.
\end{eqnarray}

\paragraph{Octupole moment}

The only nonvanishing components of the octupole moment are 
\begin{eqnarray}
Q_{33} \;=\; -Q_{3,-3} \;=\; -\sqrt{\frac{35}{64\pi}} 
\left( 6\beta_{1}^{3} \right) e
\end{eqnarray}
with
\begin{equation}
B(E3;0^{+}\rightarrow 3^{-}) \;=\; Q_{33}^{\ast} Q_{3,-3} + Q_{3,-3}^{\ast} Q_{33} ~.
\end{equation}

\subsubsection{Charge radius}

The r.m.s. charge radius can be calculated 
from the derivative of the elastic form factor as 
\ba
\left\langle r^{2}\right\rangle &=& -6 \left. \frac{d^2 F(0^+ \rightarrow 0^+;q)}{dq^2} \right|_{q=0} 
\nonumber\\
&=&\frac{3}{5}\left( \beta _{1}^{2}+\frac{3}{2\alpha _{1}}\right) +\frac{2}{5%
}\left( \beta _{2}^{2}+\frac{3}{2\alpha _{2}}\right) 
\nonumber\\
&=&\frac{3}{5}\left\langle r^{2}(\Delta) \right\rangle +\frac{2}{%
5}\left\langle r^{2}(-)\right\rangle ~. 
\label{radius}
\ea 

\section{${\cal D}_{3h}$ symmetry in $^{20}$Ne}

In order to determine the parameters of the ${\cal D}_{3h}$ configuration in 
$^{20}$Ne, we use the value of $\alpha_{1}=\alpha_{2}=0.53$ fm$^{-2}$ obtained
from the r.m.s. radius $\left\langle r^{2}\right\rangle_{\alpha}^{1/2}=1.674 \pm 0.012$ fm of the free $\alpha $-particle \cite{sick}. We also fix the values of 
$\beta_{1}$ \ and $\beta_{2}$ to have the minimum of the elastic form
factor at $q_{\min }=1.5$ fm$^{-1}$ \cite{abgrall} obtaining $\beta_{1}=1.80$ fm and $\beta_{2}=3.00$ fm. The value of $\beta_{1}$ is consistent with our
previous studies of $^{8}$Be, $^{12}$C and $^{16}$O where $\beta_{1} \sim
1.82$ fm \cite{bijker-review}. With these values we can now calculate all
quantities of interest.

\subsection{Electromagnetic transitions and moments}

With the values of $\beta_{1}=1.80$ fm and $\beta_{2}=3.00$ fm we calculate
the value of the intrinsic quadrupole moment to be $Q_{0}=52.5$ efm$^{2}$. From
this value we can calculate all $E2$ transition rates and quadrupole moments
in the ground state rotational band, assuming rigid rotations of the ${\cal D}_{3h}$
structure
\begin{eqnarray}
B(E2;L-2 \rightarrow L) &=& \frac{5}{16\pi} \, Q_{0}^{2} \, 
\left\langle L-2,0,2,0|L,0\right\rangle^{2} ~,
\nonumber\\
Q(L) &=& Q_{0} \, \left\langle L,0,2,0|L,0 \right\rangle 
\left\langle L,L,2,0|L,L \right\rangle ~.
\end{eqnarray}
In particular we obtain $B(E2;0^+ \rightarrow 2^+)=275$ e$^{2}$fm$^{4}$ and $Q(2^+)=-15.0$ efm$^{2}$ to be compared with the experimental values $327(16)$ e$^{2}$fm$^{4}$ and $-23(3)$ efm$^{2}$, respectively. The inconsistency between these two experimental values has been noted by several authors and it is common to both the cluster model
and the collective quadrupole model. We can also calculate the additional $B(E2)$ values along the ground state band given in Table~\ref{BE2}. This table indicates that the rotation of the ${\cal D}_{3h}$ structure in $^{20}$Ne 
is not rigid and that the rotational band appears to loose most of its
collectivity at $L=6$. This problem is common to the collective quadrupole model.

\begin{table}
\centering
\caption{Comparison between calculated and experimental 
$B(E2)$ values in e$^{2}$fm$^{4}$.}
\vspace{10pt}
\label{BE2}
\begin{tabular}{ccc}
\hline
\noalign{\smallskip}
& Calc & Exp \cite{tilley} \\
\noalign{\smallskip}
\hline
\noalign{\smallskip}
$B(E2;2^+ \rightarrow 0^+)$ & 55.0 & 65.4(32) \\ 
$B(E2;4^+ \rightarrow 2^+)$ & 78.4 & 70.9(64) \\ 
$B(E2;6^+ \rightarrow 4^+)$ & 86.3 & 64.5(10) \\ 
$B(E2;8^+ \rightarrow 6^+)$ & 90.4 & 29.0(42) \\
\noalign{\smallskip}
\hline
\end{tabular}
\end{table}

With the values of $\beta_1$ and $\beta_2$ given above one can also calculate hexadecapole transitions and moments. In particular, we obtain $B(E4;0^+ \rightarrow 4^+)=8.6 \times 10^{4}$ e$^{2}$fm$^{8}$ for the hexadecapole transition and $H_4=347$ efm$^{4}$ for the hexadecapole moment, in comparison with the experimental value $H_4=220(30)$ efm$^{4}$ \cite{horikawa}. The large value of the measured and
calculated hexadecapole moment is an indication of clustering.
Finally, the octupole transition is calculated to be $B(E3;0^+ \rightarrow 3^-)=210.3$ e$^{2}$fm$^{6}$. There is no experimental information on $B(E4)$ and $B(E3)$ values. 

\subsection{Radius}

The r.m.s. radius, calculated according to Eq.~(\ref{radius}), is $\left\langle
r^{2}\right\rangle^{1/2}=2.89$ fm in good agreement with the
experimental value of $3.004(25)$ fm \cite{tilley}.

\subsection{Form factors}

\begin{figure}
\centering
\begin{minipage}{0.5\linewidth}
\includegraphics[scale=0.8]{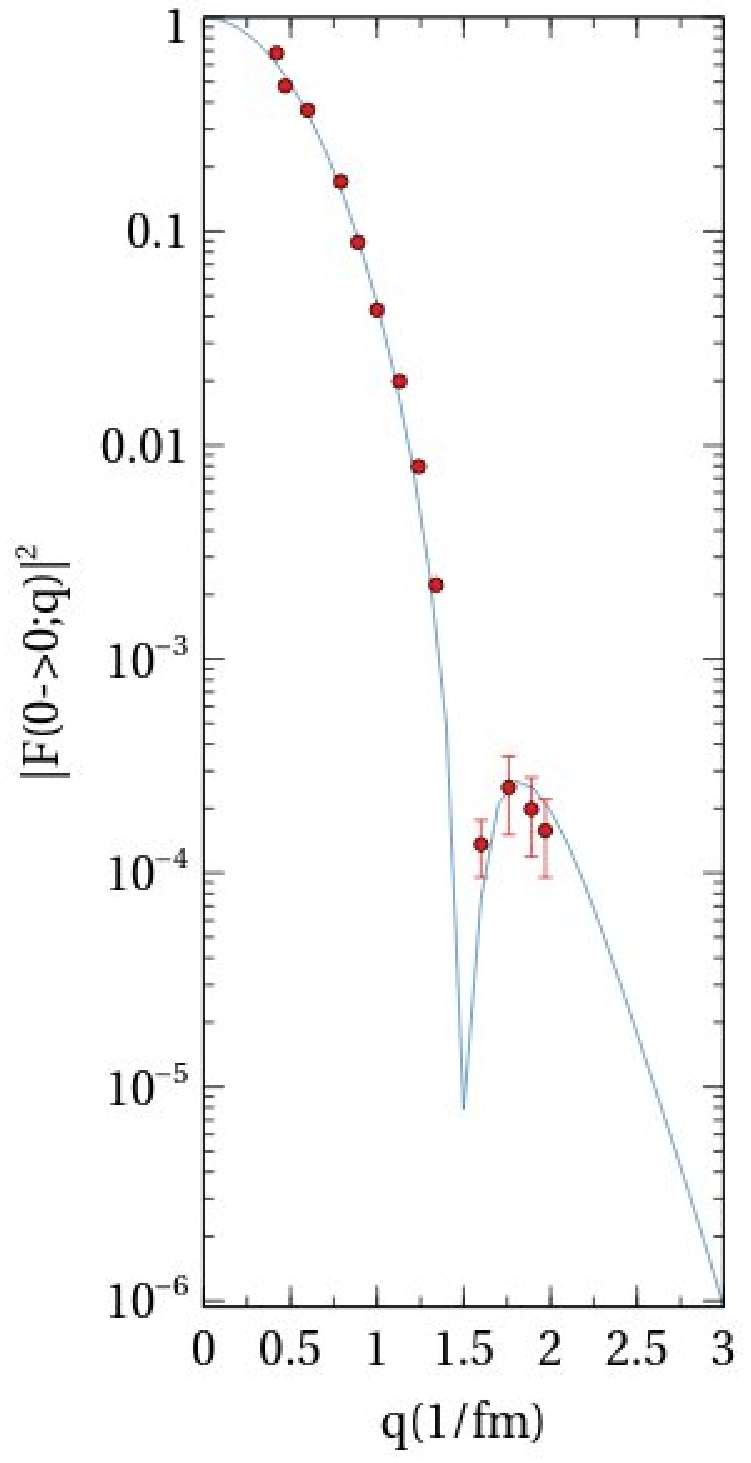}
\end{minipage}\hfill
\begin{minipage}{0.5\linewidth}
\includegraphics[scale=0.8]{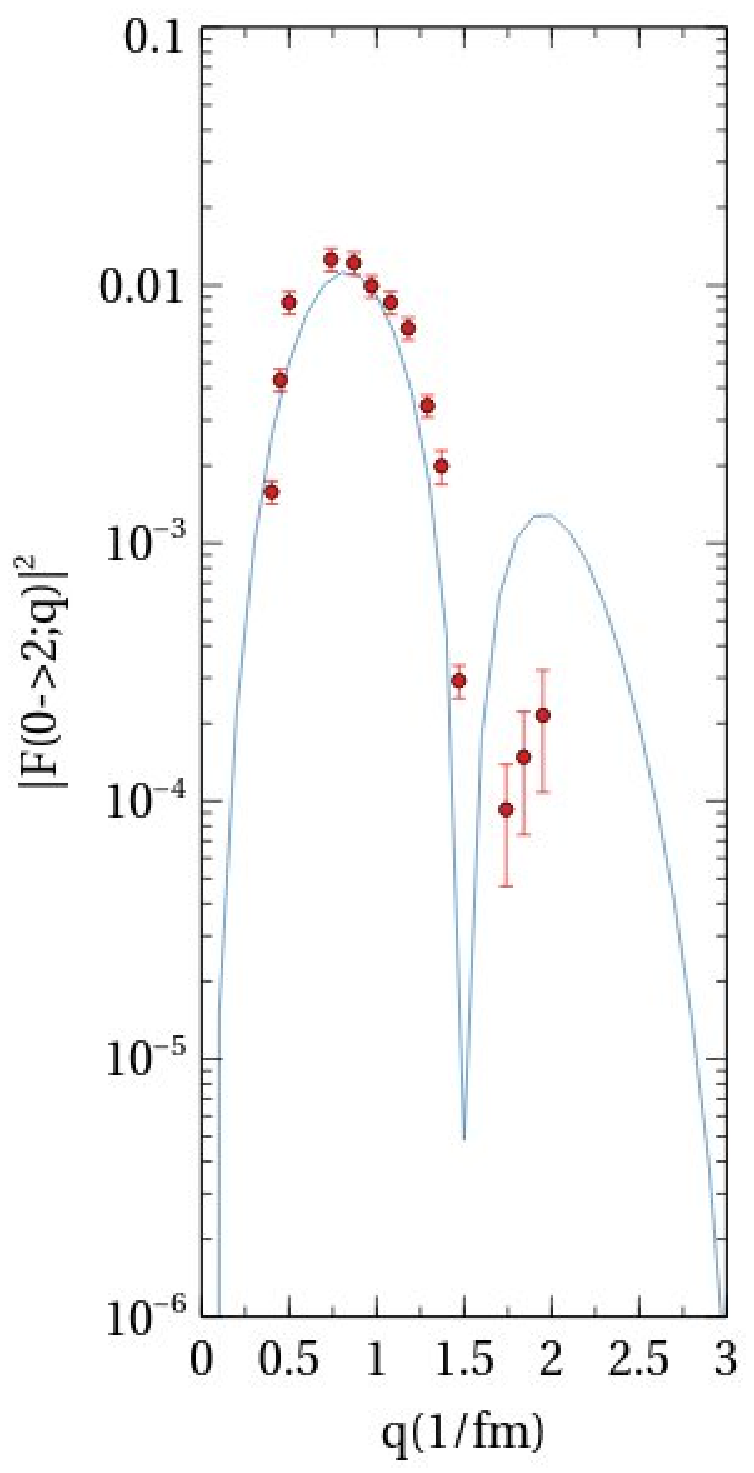}
\end{minipage}
\caption{Comparison between the experimental form factors $|F(0^+ \rightarrow L^P;q)|^2$ of $^{20}$Ne for the final states $L^P=0^+$ (elastic) and $L^P=2^+$ (1.63 MeV), and those obtained assuming bi-pyramidal ${\cal D}_{3h}$ symmetry, Eq.~(\ref{formfactors}). The experimental points are taken from Refs.~\cite{horikawa,horikawa2,abgrall}.} 
\label{ff1}
\end{figure}

\begin{figure}
\centering
\begin{minipage}{0.5\linewidth}
\includegraphics[scale=0.8]{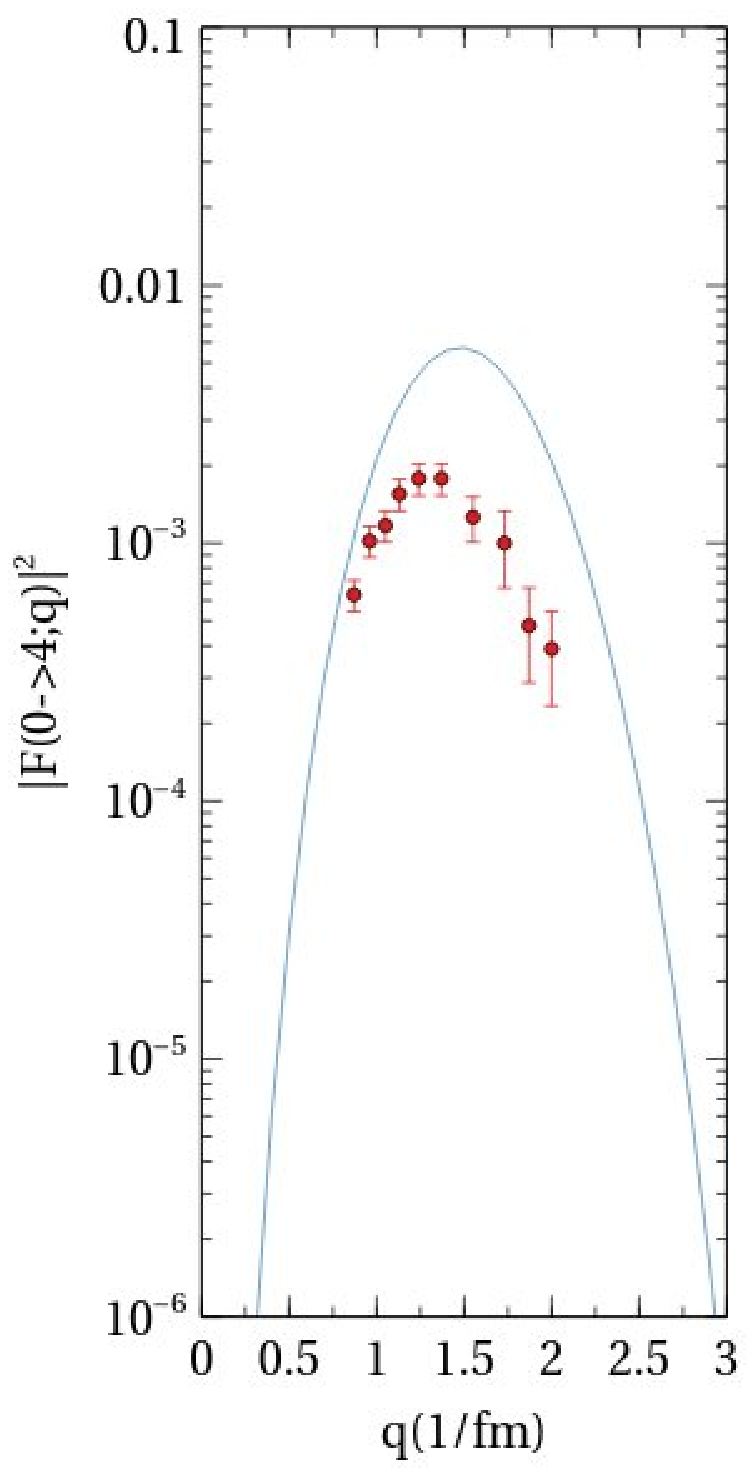}
\end{minipage}\hfill
\begin{minipage}{0.5\linewidth}
\includegraphics[scale=0.8]{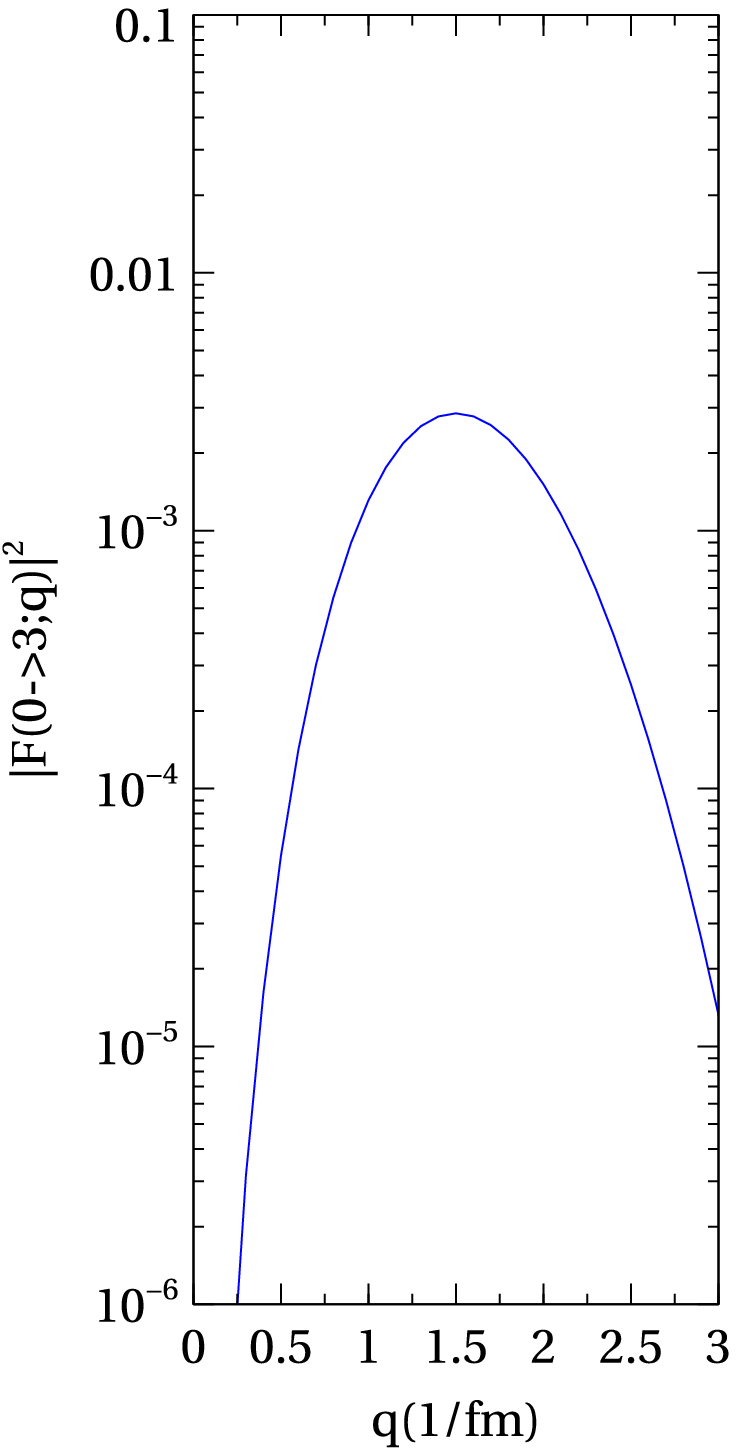}
\end{minipage}
\caption{As Fig.~\ref{ff1}, but for the final states $L^P=4^+$ (4.25 MeV) and $L^P=3^-$ (9.11 MeV).} 
\label{ff2}
\end{figure}

The calculated form factors according to Eq.~(\ref{formfactors}) are shown in Figs.~\ref{ff1} and \ref{ff2}, where they are compared with experiment \cite{horikawa,horikawa2,abgrall}. The calculated elastic form factor, 
$0^+ \rightarrow 0^+$, is in excellent agreement with experiment, the transition form factor, $0^+ \rightarrow 2^+$, in good agreement, while the transition form factor, $0^+ \rightarrow 4^+$, has the correct shape but overestimates the experiment by a factor of 2. No data are available for the transition form factor, 
$0^+ \rightarrow 3^-$. The agreement of the analytic expression of Eq.~(\ref{formfactors}) with experiment is remarkable since it shows that the bi-pyramidal configuration reproduces to high accuracy the charge density of $^{20}$Ne thus providing strong evidence for the cluster structure of this nucleus.   

\subsection{Moments of inertia}

The moments of inertia, $I$, and inertial parameters, $B$, calculated with
Eqs.~(\ref{inertia}) and (\ref{coefb}) are
\ba
\frac{I_{z}}{m} \;=\; 76.6 \mbox{ fm}^{2} ~, &\hspace{1cm}&  
\frac{I_{x}}{m} \;=\; \frac{I_{y}}{m} \;=\; 129.2 \mbox{ fm}^{2} ~,
\nonumber\\  
B_{z} \;=\; 0.270 \mbox{ MeV} ~, && B_{x} \;=\; B_{y} \;=\; 0.160 \mbox{ MeV} ~,
\ea
to be compared with the experimental value $B_{x}=B_{y}=0.212$ MeV (from the $%
4-0$ energy difference). One can also calculate the difference $%
B_{z}-B_{x}=+0.110$ MeV to be compared with that extracted by assigning the $%
3^{-}$ level at $9.11$ MeV to the ground state band with ${\cal D}_{3h}$ irrep $%
A_{1}^{\prime }$ (see Sect.~\ref{spectrum}), $B_{z}-B_{x}=+0.730$ MeV. Although the values so obtained are of the correct order of magnitude they differ considerably
from the experimental values. They however confirm the prolate nature of $^{20}$Ne, $I_{z}<I_{x}$ or $B_{z}>B_{x}$. It is interesting to note that the
values of the moments of inertia for point particles are 
\ba
\frac{I_{z}}{m} \;=\; 38.9 \mbox{ fm}^{2} ~, &\hspace{1cm}&  
\frac{I_{x}}{m} \;=\; \frac{I_{y}}{m} \;=\; 91.4 \mbox{ fm}^{2} ~, 
\nonumber\\  
B_{z} \;=\; 0.532 \mbox{ MeV} ~, && B_{x} \;=\; B_{y} \;=\; 0.226 \mbox{ MeV} ~,
\ea
and $B_{z}-B_{x}=+0.306$ MeV, in much better agreement with experiment. This
may indicate that the rotational motion is not rigid and most importantly
that the classical calculation may not be appropriate for the moment of
inertia, much in the same way as in the case of the collective model \cite%
{rowe}. The difference between the calculated $I_{x}/m=I_{y}/m=129.1$ fm$^{2}$
rigid moment of inertia and the experimental value $I_{x}/m=I_{y}/m=97.6$ fm$^{2}$ can be taken into account by introducing a quenching factor $\eta =I_{\rm exp}/I_{\rm rigid}=0.76$.

\section{Comparison with other models}
\label{emgs}

\begin{table}
\centering
\caption{Ground state properties of $^{20}$Ne.}
\label{gs}
\vspace{10pt}
\begin{tabular}{cccc}
\hline
\noalign{\smallskip}
& $\left\vert E_{0} \right\vert$ (MeV) 
& $\left\langle r^{2} \right\rangle^{1/2}$ (fm) 
& $\left\vert Q_{0} \right\vert$ (efm$^{2}$) \\ 
\noalign{\smallskip}
\hline
\noalign{\smallskip}
Exp \cite{tilley} & 160.6 & 3.004(25) 
& $\begin{array}{c} 80.5(10.5) \\ 57.3(12.7) \end{array}$ \\ 
\noalign{\smallskip}
\mbox{Brink-Bloch} \cite{brink2} & 112.9 & 2.95 & 99 \\ 
\mbox{HF} \cite{brink2} & 110.1 & 3.05 & 54 \\ 
\noalign{\smallskip}
\mbox{Bi-pyramid} ${\cal D}_{3h}$ & - & 2.89 & 52.5 \\
\noalign{\smallskip}
\hline
\end{tabular}
\end{table}

Ground state properties of $^{20}$Ne have been calculated in a variety of
models. We begin by considering the binding energy, $\left\vert
E_{0} \right\vert$, r.m.s. radius, $\left\langle r^{2}\right\rangle^{1/2}$,
and intrinsic quadrupole moment, $\left\vert Q_{0} \right\vert $. These are
given in Table~\ref{gs}.
We note that the calculated values in the Brink-Bloch model and in
Hartree-Fock depend on the force used in the calculation. The values given
in Table~\ref{gs} are for the Brink-Boeker force \cite{brink-boeker}. The two
experimental values of $\left\vert Q_{0} \right\vert $ are from the
quadrupole moment $Q(2^{+})$ and the $B(E2;2^{+}\rightarrow 0^{+})$
respectively. The inconsistency between these two experimental values has
been noted by several authors, and calls for a new measurement.

\begin{table}
\centering
\caption{Energy levels of the ground state band in $^{20}$Ne in MeV.}
\label{gsb}
\vspace{10pt}
\begin{tabular}{ccccc}
\hline
\noalign{\smallskip}
$L$ & $E_{\rm exp}$ \cite{tilley} & $E_{{\cal D}_{3h}}$ 
& $E_{\rm BB}$ \cite{weiguny1} & $E_{\rm SM}$ \cite{wong} \\ 
\noalign{\smallskip}
\hline
\noalign{\smallskip} 
0 & 0.00 & 0.00 & 0.00 & 0.00 \\ 
2 & 1.63 & 1.26 & 1.24 & 1.66 \\ 
4 & 4.25 & 4.20 & 3.87 & 3.94 \\ 
6 & 8.78 & 8.82$^{\ast}$ & 8.79$^{\ast}$ & 8.28 \\ 
8 & (11.95) & 15.12 &  & 11.48 \\
\noalign{\smallskip}
\hline
\end{tabular}
\end{table}

Next we consider the energy levels of the ground state band in $^{20}$Ne.
These are given in Table~\ref{gsb}.
The star in the Brink-Bloch calculation indicates that this value has been
used to choose the force. The star in the ${\cal D}_{3h}$ calculation indicates
that this value has been used to determine the quenching parameter $\eta=0.76$. 
The shell model calculation is a $sd$ shell calculation as reported
on p. 268 of Ref.~\cite{wong}. From this table one can see that while the cluster
model describes reasonably well states up to $6^{+}$, it does not describe
well the energy of the $8^{+}$ state. The shell model instead describes very
well the observed energies. The same situation occurs for the B(E2) values,
as shown in Table~\ref{E2}.
One can see that the $8^{+}$ state does not appear to be a member of the
rotational band, and that the shell model underestimates the experimental
values even if large effective charges, $e_{n}=0.5e$ and $e_{p}=1.5e$, are
used. It gives however the correct trend including the $8^{+}\rightarrow
6^{+}$ transition.

\begin{table}
\centering
\caption{$B(E2)$ values for $L \rightarrow L-2$ transitions along the 
ground-state band of $^{20}$Ne in e$^{2}$fm$^{4}$.}
\label{E2}
\vspace{10pt}
\begin{tabular}{cccc}
\hline
\noalign{\smallskip}
$L$ & Exp & ${\cal D}_{3h}$ & SM \\
\noalign{\smallskip}
\hline
\noalign{\smallskip}
$2$ & $65.4 \pm 3.2$ & 55 & 48 \\ 
$4$ & $70.9 \pm 6.4$ & 78 & 58 \\ 
$6$ & $64.5 \pm 1.0$ & 86 & 43 \\ 
$8$ & $29.0 \pm 4.2$ & 90 & 28 \\
\noalign{\smallskip}
\hline
\end{tabular}
\end{table}

In summary, the cluster model with ${\cal D}_{3h}$ symmetry and rigid rotations
describes ground state experimental data at the same level of the
Brink-Bloch and HF models, but not as well as the shell model. However,
ground state properties are only a test of collective versus shell model and
not of the specific nature of the collective motion. In order to distinguish
between different collective models, one must analyze the excitation
spectrum to which we now revert.

\section{Excitation spectrum of $^{20}$Ne: Evidence for ${\cal D}_{3h}$ symmetry}
\label{spectrum}

The excitation spectrum of a bi-pyramidal configuration with ${\cal D}_{3h}$
symmetry can be analyzed using standard methods of molecular physics. We
begin by considering the representations of the point group ${\cal D}_{3h}$ denoted
generically by $\Gamma$. Each representation $\Gamma $ of ${\cal D}_{3h}$ contains
a discrete (but infinite) number of values of $K^{P}$, where $K$ is the
projection of the angular momentum on the $z$-axis of Fig.~\ref{bipyramid}, and $P$ the
parity. The results are given in Table~\ref{gamma}, where both the notation used in
crystal physics \cite{koster} and in molecular physics \cite{herzberg} is
given. In this article, we use the more familiar molecular physics notation.
The representations with $'$ are related to those with $''$ by a change in parity, $P \rightarrow -P$. On top of each value of $K$, there is a rotational band with 
\begin{equation}
L \;=\; K, K+1, K+2, \ldots ~,
\end{equation}
except for $K=0$ for which the values of $L$ are even or odd as indicated in
Table~\ref{gamma}.
Note that only the representations $\Gamma_{1} \equiv A'_{1}$, 
$\Gamma_{4} \equiv A''_{2}$, $\Gamma_{5} \equiv E'$, and $\Gamma _{6} \equiv E''$ appear in this paper which deals with vibrational (bosonic) states.

\begin{table}
\centering
\caption{Values of $K^{P}$ contained in each representation $\Gamma$ 
of ${\cal D}_{3h}$.}
\label{gamma}
\vspace{10pt}
\begin{tabular}{ccl}
\hline
\noalign{\smallskip}
$\Gamma$ \cite{koster} & $\Gamma$ \cite{herzberg} & $K^{P}$ bands \\ 
\noalign{\smallskip}
\hline
\noalign{\smallskip}
$\Gamma_{1}$ & $A'_{1}$  & $0^{+}(L=\mbox{even})$, $3^{-}$, $6^{+}$, $\ldots$ \\ 
$\Gamma_{2}$ & $A''_{1}$ & $0^{-}(L=\mbox{even})$, $3^{+}$, $6^{-}$, $\ldots$ \\ 
$\Gamma_{3}$ & $A'_{2}$  & $0^{+}(L=\mbox{ odd})$, $3^{-}$, $6^{+}$, $\ldots$ \\ 
$\Gamma_{4}$ & $A''_{2}$ & $0^{-}(L=\mbox{ odd})$, $3^{+}$, $6^{-}$, $\ldots$ \\ 
$\Gamma_{5}$ & $E'$      & $1^{-}$, $2^{+}$, $4^{+}$, $5^{-}$, $\ldots$ \\ 
$\Gamma_{6}$ & $E''$     & $1^{+}$, $2^{-}$, $4^{-}$, $5^{+}$, $\ldots$ \\
\noalign{\smallskip}
\hline
\end{tabular}
\end{table}

\subsection{Representations of a bi-pyramidal configuration}

\begin{figure}
\centering
\vspace{15pt}
\setlength{\unitlength}{0.9pt}
\begin{picture}(360,520)(0,0)
\thicklines
\put( 75,425) {\circle*{8}} 
\put( 30,440) {\circle*{8}}
\put( 90,440) {\circle*{8}}
\put( 60,500) {\circle*{8}}
\put( 60,380) {\circle*{8}}
\put(  0,370) {$\nu_2(A'_1)$}
\put( 60,500) {\vector( 0, 1){20}}
\put( 60,380) {\vector( 0,-1){20}}
\multiput( 30,440)(4,0){15}{\circle*{1}}
\put( 30,440) {\line( 1, 2){30}}
\put( 30,440) {\line( 3,-1){45}}
\put( 30,440) {\line( 1,-2){30}}
\put( 90,440) {\line(-1, 2){30}}
\put( 90,440) {\line(-1,-2){30}}
\put( 90,440) {\line(-1,-1){15}}
\put( 75,425) {\line(-1, 5){15}}
\put( 75,425) {\line(-1,-3){15}}
\put(195,425) {\circle*{8}} 
\put(150,440) {\circle*{8}}
\put(210,440) {\circle*{8}}
\put(180,500) {\circle*{8}}
\put(180,380) {\circle*{8}}
\put(120,370) {$\nu_{6a}(E'')$}
\put(195,425) {\vector( 0, 1){28}} 
\put(150,440) {\vector( 0,-1){14}}
\put(210,440) {\vector( 0,-1){14}}
\put(180,500) {\vector( 2,-3){ 8}}
\put(180,380) {\vector(-2, 3){ 8}}
\multiput(150,440)(4,0){15}{\circle*{1}}
\put(150,440) {\line( 1, 2){30}}
\put(150,440) {\line( 3,-1){45}}
\put(150,440) {\line( 1,-2){30}}
\put(210,440) {\line(-1, 2){30}}
\put(210,440) {\line(-1,-2){30}}
\put(210,440) {\line(-1,-1){15}}
\put(195,425) {\line(-1, 5){15}}
\put(195,425) {\line(-1,-3){15}}
\put(315,425) {\circle*{8}} 
\put(270,440) {\circle*{8}}
\put(330,440) {\circle*{8}}
\put(300,500) {\circle*{8}}
\put(300,380) {\circle*{8}}
\put(240,370) {$\nu_{6b}(E'')$}
\put(270,440) {\vector( 0,-1){24}}
\put(330,440) {\vector( 0, 1){24}}
\put(300,500) {\vector( 1, 0){14}}
\put(300,380) {\vector(-1, 0){14}}
\multiput(270,440)(4,0){15}{\circle*{1}}
\put(270,440) {\line( 1, 2){30}}
\put(270,440) {\line( 3,-1){45}}
\put(270,440) {\line( 1,-2){30}}
\put(330,440) {\line(-1, 2){30}}
\put(330,440) {\line(-1,-2){30}}
\put(330,440) {\line(-1,-1){15}}
\put(315,425) {\line(-1, 5){15}}
\put(315,425) {\line(-1,-3){15}}
\put( 75,255) {\circle*{8}} 
\put( 30,270) {\circle*{8}}
\put( 90,270) {\circle*{8}}
\put( 60,330) {\circle*{8}}
\put( 60,210) {\circle*{8}}
\put(  0,200) {$\nu_3(A'_1)$}
\put( 75,255) {\vector( 2,-3){12}} 
\put( 30,270) {\vector(-4, 1){20}}
\put( 90,270) {\vector( 2, 1){18}}
\multiput( 30,270)(4,0){15}{\circle*{1}}
\put( 30,270) {\line( 1, 2){30}}
\put( 30,270) {\line( 3,-1){45}}
\put( 30,270) {\line( 1,-2){30}}
\put( 90,270) {\line(-1, 2){30}}
\put( 90,270) {\line(-1,-2){30}}
\put( 90,270) {\line(-1,-1){15}}
\put( 75,255) {\line(-1, 5){15}}
\put( 75,255) {\line(-1,-3){15}}
\put(195,255) {\circle*{8}} 
\put(150,270) {\circle*{8}}
\put(210,270) {\circle*{8}}
\put(180,330) {\circle*{8}}
\put(180,210) {\circle*{8}}
\put(120,200) {$\nu_{4a}(E')$}
\put(195,255) {\vector( 2,-3){12}} 
\put(150,270) {\vector( 2, 1){18}}
\put(210,270) {\vector(-2, 1){18}}
\multiput(150,270)(4,0){15}{\circle*{1}}
\put(150,270) {\line( 1, 2){30}}
\put(150,270) {\line( 3,-1){45}}
\put(150,270) {\line( 1,-2){30}}
\put(210,270) {\line(-1, 2){30}}
\put(210,270) {\line(-1,-2){30}}
\put(210,270) {\line(-1,-1){15}}
\put(195,255) {\line(-1, 5){15}}
\put(195,255) {\line(-1,-3){15}}
\put(315,255) {\circle*{8}} 
\put(270,270) {\circle*{8}}
\put(330,270) {\circle*{8}}
\put(300,330) {\circle*{8}}
\put(300,210) {\circle*{8}}
\put(240,200) {$\nu_{4b}(E')$}
\put(315,255) {\vector(-1, 0){21}} 
\put(270,270) {\vector( 3,-1){20}}
\put(330,270) {\vector( 1, 1){12}}
\multiput(270,270)(4,0){15}{\circle*{1}}
\put(270,270) {\line( 1, 2){30}}
\put(270,270) {\line( 3,-1){45}}
\put(270,270) {\line( 1,-2){30}}
\put(330,270) {\line(-1, 2){30}}
\put(330,270) {\line(-1,-2){30}}
\put(330,270) {\line(-1,-1){15}}
\put(315,255) {\line(-1, 5){15}}
\put(315,255) {\line(-1,-3){15}}
\put( 75, 85) {\circle*{8}} 
\put( 30,100) {\circle*{8}}
\put( 90,100) {\circle*{8}}
\put( 60,160) {\circle*{8}}
\put( 60, 40) {\circle*{8}}
\put(  0, 30) {$\nu_1(A''_2)$}
\put( 75, 85) {\vector( 0,-1){14}} 
\put( 30,100) {\vector( 0,-1){14}}
\put( 90,100) {\vector( 0,-1){14}}
\put( 60,160) {\vector( 0, 1){21}}
\put( 60, 40) {\vector( 0, 1){21}}
\multiput( 30,100)(4,0){15}{\circle*{1}}
\put( 30,100) {\line( 1, 2){30}}
\put( 30,100) {\line( 3,-1){45}}
\put( 30,100) {\line( 1,-2){30}}
\put( 90,100) {\line(-1, 2){30}}
\put( 90,100) {\line(-1,-2){30}}
\put( 90,100) {\line(-1,-1){15}}
\put( 75, 85) {\line(-1, 5){15}}
\put( 75, 85) {\line(-1,-3){15}}
\put(195, 85) {\circle*{8}} 
\put(150,100) {\circle*{8}}
\put(210,100) {\circle*{8}}
\put(180,160) {\circle*{8}}
\put(180, 40) {\circle*{8}}
\put(120, 30) {$\nu_{5a}(E')$}
\put(195, 85) {\vector( 2,-3){ 8}} 
\put(150,100) {\vector( 2,-3){ 8}}
\put(210,100) {\vector( 2,-3){ 8}}
\put(180,160) {\vector(-2, 3){12}}
\put(180, 40) {\vector(-2, 3){12}}
\multiput(150,100)(4,0){15}{\circle*{1}}
\put(150,100) {\line( 1, 2){30}}
\put(150,100) {\line( 3,-1){45}}
\put(150,100) {\line( 1,-2){30}}
\put(210,100) {\line(-1, 2){30}}
\put(210,100) {\line(-1,-2){30}}
\put(210,100) {\line(-1,-1){15}}
\put(195, 85) {\line(-1, 5){15}}
\put(195, 85) {\line(-1,-3){15}}
\put(315, 85) {\circle*{8}} 
\put(270,100) {\circle*{8}}
\put(330,100) {\circle*{8}}
\put(300,160) {\circle*{8}}
\put(300, 40) {\circle*{8}}
\put(240, 30) {$\nu_{5b}(E')$}
\put(315, 85) {\vector(-1,0){14}} 
\put(270,100) {\vector(-1,0){14}}
\put(330,100) {\vector(-1,0){14}}
\put(300,160) {\vector( 1,0){21}}
\put(300, 40) {\vector( 1,0){21}}
\multiput(270,100)(4,0){15}{\circle*{1}}
\put(270,100) {\line( 1, 2){30}}
\put(270,100) {\line( 3,-1){45}}
\put(270,100) {\line( 1,-2){30}}
\put(330,100) {\line(-1, 2){30}}
\put(330,100) {\line(-1,-2){30}}
\put(330,100) {\line(-1,-1){15}}
\put(315, 85) {\line(-1, 5){15}}
\put(315, 85) {\line(-1,-3){15}}
\end{picture}
\caption{Normal vibrations of a bipyramidal configuration with ${\cal D}_{3h}$ symmetry. Vertical arrows correspond to oscillations in the direction of the symmetry axis ($z$-axis), all others to oscillations in directions perpendicular to the symmetry axis.}
\label{modes}
\end{figure}
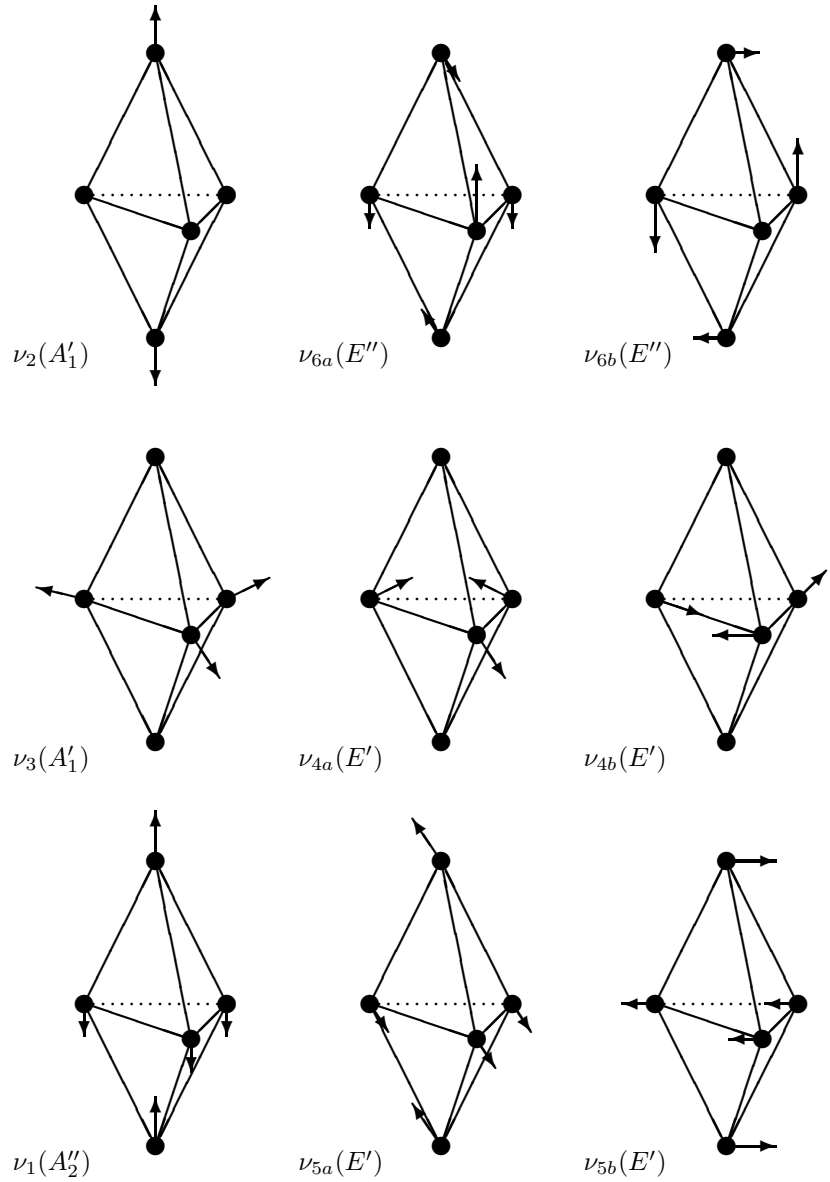

The ground state representation of a bi-pyramidal configuration with
symmetry ${\cal D}_{3h}$ is the totally symmetric representation $A_{1}^{\prime }$.
For a bi-pyramidal configuration of Fig.~\ref{bipyramid} with five particles, three of which are in the $xy$-plane and two on opposite ends along the $z$-axis, there are
$15-6=9$ vibrations, three singly degenerate and three doubly degenerate \cite%
{herzberg} (see Fig.~\ref{modes}). The species of these vibrations, their characterization and numbering are: (i) stretching vibrations of the two $\alpha$-particles on the $z$-axis with representations $A_{2}^{\prime \prime}$ (antisymmetric stretching) and $A_{1}^{\prime }$ (symmetric stretching) with energies $\omega_{1}$ and $\omega_{2}$; (ii) vibrations of the triangle \cite{bijker2} in the $xy$-plane with representations $A_{1}^{\prime}$ (stretching) and $E^{\prime}$ (bending) with energies $\omega_{3}$ and $\omega_{4}$; (iii) bending and twisting 
vibrations of the two $\alpha$-particles on the $z$-axis with representations, $E^{\prime}$ and $E^{\prime \prime}$, and energies, $\omega_{5}$ and $\omega_{6}$, respectively. 

The energy levels of a bi-pyramidal structure with ${\cal D}_{3h}$ symmetry can
then be written as those of a symmetric top (rotational part) and in the
harmonic approximation (vibrational part)
\begin{equation}
E\left( \left[ v\right] ,K,L\right) \;=\; E_{0} + \sum_{i=1}^{6}\omega _{i}v_{i} 
+ B_{x\left[ v\right] }L(L+1) + \left[B_{z}-B_{x}\right] _{\left[ v\right] }K^{2}~,
\label{energies}
\end{equation}
where $\left[ v\right] \equiv \left[ v_{1},v_{2},v_{3},v_{4},v_{5},v_{6} \right]$ is the vibrational quantum number, and $B=\hbar ^{2}/2I$ are the rotational constants assuming the frame of reference as in Fig.~\ref{bipyramid}. In this notation, the ground state is $\left[ 0\right] \equiv \left[ 0,0,0,0,0,0\right]$. The zero-point energies are included in $E_{0}$ of Eq.~(\ref{energies}).

\subsection{Assignment of states to bands and their ${\cal D}_{3h}$ classification}

The excitation spectrum of $^{20}$Ne has been extensively investigated \cite%
{tilley} up to excitation energies of $\sim 20$ MeV. An assignment of some
of the states to $K^{P}$ bands has been given on page 308 of Ref.~\cite{tilley}. 
Here we extend the assignment to additional bands, extract the rotational $B_{\left[
v\right]}$ and vibrational $\omega_{\left[ v \right] }$ parameters, and
classify them in terms of representations of ${\cal D}_{3h}$. Bands are identified
by their ${\cal D}_{3h}$ representation, their $K^{P}$ value, and the energy of the
lowest state in the band. We note that assignments of states with energy $>10$ MeV to bands is difficult due to the high density of states of the
same spin and parity above this energy and to the possible occurrence of
non-collective states. We use, in addition to energies, also electromagnetic
decay properties and intensities of transfer reactions, when available. Our
assignments, together with those of \cite{tilley} when available, for the
ground state band and the nine expected vibrational bands are given in 
Table~\ref{rotbands}. Comments for each band are given in the following subsections.

\begin{table}
\centering
\caption{Assignments of states into rotational bands.}
\label{rotbands}
\vspace{10pt}
\begin{tabular}{ccccccc}
\hline
\noalign{\smallskip}
$[v_{1},v_{2},v_{3},v_{4},v_{5},v_{6}]$ & $\Gamma$ & $K^{P}$ & $L^{P}$ 
& $E_{\rm exp}$ & $E_{\rm th}$ & $E_{\rm exp}$ \cite{tilley} \\ 
\noalign{\smallskip}
\hline
\noalign{\smallskip}
$[0,0,0,0,0,0]$ & $A'_{1}$ & $0^{+}(0.00)$ 
    & $0^{+}$ &  0.00 &  0.00 &  0.00 \\ 
& & & $2^{+}$ &  1.63 &  1.27 &  1.63 \\ 
& & & $4^{+}$ &  4.25 &  4.24 &  4.25 \\ 
& & & $6^{+}$ &  8.78 &  8.90 &  8.78 \\ 
& & & $8^{+}$ & 11.95 & 15.26 & 11.95 \\ 
\noalign{\smallskip} 
$[1,0,0,0,0,0]$ & $A''_{2}$ & $0^{-}(5.79)$ 
    & $1^{-}$ &  5.79 &  5.79 &   5.79 \\ 
& & & $3^{-}$ &  7.16 &  7.16 &   7.16 \\ 
& & & $5^{-}$ & 10.26 &  9.63 &  10.26 \\ 
& & & $7^{-}$ & 13.69 & 13.19 &  13.69 \\ 
& & & $9^{-}$ & 17.43 & 17.85 & (17.43) \\ 
\noalign{\smallskip}
$[0,1,0,0,0,0]$ & $A'_{1}$ & $0^{+}(6.72)$ 
    & $0^{+}$ &  6.72 &  6.72 &  6.73 \\ 
& & & $2^{+}$ &  7.42 &  7.48 &  7.42 \\ 
& & & $4^{+}$ &  9.03 &  9.26 &  9.99 \\ 
& & & $6^{+}$ & 12.14 & 12.05 & (12.59, 13.11) \\ 
& & & $8^{+}$ & 15.87 & 15.87 &  \\ 
\noalign{\smallskip}
$[0,0,1,0,0,0]$ & $A'_{1}$ & $0^{+}(7.19)$ 
    & $0^{+}$ &  7.19 &  7.19 &  7.20 \\ 
& & & $2^{+}$ &  7.83 &  7.97 &  7.83 \\ 
& & & $4^{+}$ &  9.99 &  9.79 &  9.03 \\ 
& & & $6^{+}$ & 12.58 & 12.65 & 12.14 \\ 
& & & $8^{+}$ & 16.75 & 16.55 &  \\ 
\noalign{\smallskip}
$[0,0,0,1,0,0]$ & $E'$ & $1^{-}(8.84)$ 
    & $1^{-}$ &   8.84  &  8.84 &  8.85 \\ 
& & & $2^{-}$ &   9.32  &  9.38 &       \\ 
& & & $3^{-}$ &  10.41  & 10.18 & 10.41 \\ 
& & & $4^{-}$ & (11.53) & 11.25 &       \\ 
& & & $5^{-}$ &  12.71  & 12.59 & 12.71 \\ 
& & & $6^{-}$ &         & 14.20 &       \\ 
& & & $7^{-}$ &  16.58  & 16.08 & 16.58 \\ 
& & & $8^{-}$ &         & 18.22 &       \\ 
& & & $9^{-}$ &  20.69  & 20.63 & (20.69, 21.06) \\ 
\noalign{\smallskip}
& & $2^{+}(9.20)$ 
    & $2^{+}$ &  9.20 &  9.20 &  \\ 
& & & $3^{+}$ &  9.87 &  9.87 &  \\ 
& & & $4^{+}$ & 10.55 & 10.77 &  \\ 
& & & $5^{+}$ &       & 11.89 &  \\ 
& & & $6^{+}$ & 13.10 & 13.23 &  \\ 
& & & $7^{+}$ &       & 14.80 &  \\ 
& & & $8^{+}$ & 17.29 & 16.59 &  \\ 
\noalign{\smallskip}
\hline
\end{tabular}
\end{table}

\addtocounter{table}{-1}
\begin{table}
\centering
\caption{Continued.}
\vspace{10pt}
\begin{tabular}{ccccccc}
\hline
\noalign{\smallskip}
$[v_{1},v_{2},v_{3},v_{4},v_{5},v_{6}]$ & $\Gamma$ & $K^{P}$ & $L^{P}$ 
& $E_{\rm exp}$ & $E_{\rm th}$ & $E_{\rm exp}$ \cite{tilley} \\ 
\noalign{\smallskip}
\hline
\noalign{\smallskip}
$[0,0,0,0,1,0]$ & $E'$ & $1^{-}(8.71)$ 
    & $1^{-}$ &   8.71  &  8.71 &  \\ 
& & & $2^{-}$ &         &  9.30 &  \\ 
& & & $3^{-}$ &  10.84  & 10.18 &  \\ 
& & & $4^{-}$ &         & 11.36 &  \\ 
& & & $5^{-}$ &  13.42  & 12.82 &  \\ 
& & & $6^{-}$ &         & 14.59 &  \\ 
& & & $7^{-}$ & (16.63) & 16.65 &  \\ 
& & & $8^{-}$ &         & 19.00 &  \\ 
& & & $9^{-}$ &  21.06  & 21.64 &  \\ 
\noalign{\smallskip}
& & $2^{+}(9.49)$ 
    & $2^{+}$ &   9.49  &  9.49 &  \\ 
& & & $3^{+}$ & (10.69) & 10.27 &  \\ 
& & & $4^{+}$ &  11.02  & 11.31 &  \\ 
& & & $5^{+}$ &         & 12.61 &  \\ 
& & & $6^{+}$ &  13.93  & 14.17 &  \\ 
& & & $7^{+}$ &         & 15.99 &  \\ 
& & & $8^{+}$ & (18.96) & 18.07 &  \\ 
\noalign{\smallskip}
$[0,0,0,0,0,1]$ & $E''$ & $1^{+}(9.93)$  
    & $1^{+}$ & (9.93) &  9.93 &  \\ 
& & & $2^{+}$ & 10.58  & 10.42 &  \\ 
& & & $3^{+}$ & 10.92  & 11.17 &  \\ 
& & & $4^{+}$ & 12.25  & 12.16 &  \\ 
& & & $5^{+}$ &        & 13.40 &  \\ 
& & & $6^{+}$ & 14.31  & 14.89 &  \\ 
& & & $7^{+}$ &        & 16.63 &  \\ 
& & & $8^{+}$ & 18.62  & 18.61 &  \\ 
\noalign{\smallskip}
& & $2^{-}(4.97)$ 
    & $2^{-}$ &   4.97  &  4.97 &   4.97  \\ 
& & & $3^{-}$ &   5.62  &  5.84 &   5.62  \\ 
& & & $4^{-}$ &   7.00  &  7.00 &   7.00  \\ 
& & & $5^{-}$ &   8.45  &  8.45 &   8.46  \\ 
& & & $6^{-}$ &  10.60  & 10.19 &  10.61  \\ 
& & & $7^{-}$ &  13.34  & 12.22 &  13.34  \\ 
& & & $8^{-}$ & (15.70) & 15.41 & (15.70) \\ 
& & & $9^{-}$ &  17.43  & 17.15 &  17.43  \\ 
\noalign{\smallskip}
\hline
\end{tabular}
\end{table}

\subsubsection{$[0,0,0,0,0,0]$ $A'_{1}$: $K^{P}=0^{+}(0.00)$}

This is the ground state band. Experimental and calculated energies
with $B=212$ keV are given in Table~\ref{rotbands} where they are compared with the 
assignments of \cite{tilley}. The energy of the $8^{+}$ state does not fit the rotational formula, nor does the $B(E2;8^{+} \rightarrow 6^{+})$ value. We believe that this state is not a collective cluster state. This band belongs to the representation $A_{1}^{\prime }$ of ${\cal D}_{3h}$. This representation contains in addition to the band with $K^{P}=0^{+}$ also a high-lying band with $K^{P}=3^{-}$. We
tentatively identify the lowest state of this band with the $3^{-}$ state at $9.11$ MeV.

\subsubsection{$[1,0,0,0,0,0]$ $A''_{2}$: $K^{P}=0^{-}(5.79)$}

This band is assigned to the representation $A''_{2}$ of ${\cal D}_{3h}$. Its interpretation is an oscillation in the $z$ direction of the two $\alpha$-particles on the $z$-axis with respect to the three $\alpha$-particles in the $xy$-plane (anti-symmetric stretching, $\nu_1$ in Fig.~\ref{modes}). The rotational and vibrational values are $B=137$ keV and $\omega=5.52$ MeV. The assignment of the $9^{-}$ state in \cite{tilley} is tentative and therefore in parenthesis, but it fits almost exactly the rotational behavior. This band has been assigned by von Oertzen \cite{vonoertzen} to the anti-symmetric oscillation of an $\alpha$-particle relative to $^{16}$O. Since the bi-pyramid can be thought of as a tetrahedron ($^{16}$O) plus an $\alpha$-particle on the $z$-axis, the interpretation of von Oertzen is similar to ours, except for the reduction of ${\cal D}_{3h}$ symmetry to $D_{3}$. 

\subsubsection{$[0,1,0,0,0,0]$ $A'_{1}$: $K^{P}=0^{+}(6.72)$}

This band is assigned to the $A_{1}^{\prime }$ representation of ${\cal D}_{3h}$ 
which is interpreted as the symmetric vibration of the two $\alpha $-particles
on the $z$-axis (symmetric stretching, $\nu_2$ in Fig.~\ref{modes}). The rotational and vibrational parameters are $B=127$ keV and $\omega =6.72$ MeV. The rotational parameter of this band $B=127$ keV is close to that of the band with $A''_{2}$ symmetry, $B=137$ keV, indicating that the geometric structure of both bands is very
similar, if not identical. A comparison with the rotational parameter of the
ground state band, $B=212$ keV, shows that the geometric structure of these
two bands is more extended than that of the ground state. In the assignment
of \cite{tilley}, two values are given in parenthesis. We opt for the first
of these values.

\subsubsection{$[0,0,1,0,0,0]$ $A'_{1}$: $K^{P}=0^{+}(7.19)$}

This band is assigned to the $A_{1}^{\prime }$ representation of ${\cal D}_{3h}$.
Its interpretation is the symmetric stretching vibration of the triangle ($\nu_3$ in Fig.~\ref{modes}). The rotational and vibrational parameters are $B=130$ keV and $\omega =7.19$ MeV.
The interpretation as stretching vibration of the triangle is supported by
its vibrational energy, $7.19$ MeV, which is remarkably close to the energy
of the stretching vibration in $^{12}$C, the so-called Hoyle state, $7.65$ MeV. Indeed, the occurrence of this band provides strong evidence for ${\cal D}_{3h}$ symmetry in $^{20}$Ne. We note in passing that because the quantum
numbers of the three bands $A_{1}^{\prime}:K^{P}=0^{+}(0.00)$, $A_{1}^{\prime}:K^{P}=0^{+}(6.72)$ and $A_{1}^{\prime}:K^{P}=0^{+}(7.19)$ are
the same, these bands may be mixed by rotation-vibration interactions.

\subsubsection{$[0,0,0,1,0,0]$ $E'$: $K^{P}=1^{-}(8.84)$ and $K^{P}=2^{+}(9.20)$}

The representation $E^{\prime }$ is doubly degenerate with the two lowest
values of $K^{P}$ being $K^{P}=1^{-}$, $2^{+}$. Its interpretation is that of a two-dimensional bending vibration of the triangle according to Fig.~4 of 
\cite{bijker-review} ($\nu_4$ in Fig.~\ref{modes}). The rotational and vibrational parameters for the $K^{P}=1^{-}$ band are $B=134$ keV and $\omega =8.59$ MeV. 
This interpretation is supported by the almost equality of its rotational parameter with that of the stretching vibration of the triangle, $B=130$ MeV. Note that this band, having $K^{P}=1^{-}$ has both odd and even values of $L$. In \cite{tilley} only the odd values are reported.

The rotational and vibrational parameters for the $K^{P}=2^{+}$ band are $B=212$ keV  and $\omega =8.53$ MeV. The almost equality of its vibrational energy with that of the $K^{P}=1^{-}$ band emphasizes that both rotational bands belong to the same ${\cal D}_{3h}$ representation $E^{\prime}$. This band is not assigned in \cite{tilley}.

\subsubsection{$[0,0,0,0,1,0]$ $E'$: $K^{P}=1^{-}(8.71)$ and $K^{P}=2^{+}(9.49)$}

We assign these bands to the two-dimensional bending vibration of the two $\alpha$-particles on the $z$-axis with $E^{\prime}$ symmetry ($\nu_5$ in Fig.~\ref{modes}). The rotational and vibrational parameters for the $K^{P}=1^{-}$ band are $B=147$ keV and $\omega=8.42$ MeV. We note that above $\sim 10$ MeV of excitation, it is difficult to assign levels to bands, due to the high density of states at this energy and to the fact that in $\alpha$ scattering experiments $^{20}$Ne($\alpha,\alpha^{\prime}$) from which most of the information comes, it is not possible to excite unnatural parity states. Our assignments here are
therefore tentative.

The rotational and vibrational parameters for the $K^{P}=2^{+}$ band are $B=130$ keV and $\omega=8.71$ MeV. The same comments as for the band $K^{P}=1^{-}$ apply here,
although there is some evidence for the unnatural parity state $J^P=3^{+}$.

\subsubsection{$[0,0,0,0,0,1]$ $E''$: $K^{P}=1^{+}(9.93)$ and $K^{P}=2^{-}(4.97)$}

The representation $E''$ is doubly degenerate with the two
lowest values of $K^{P}$ being $K^{P}=1^{+}$, $2^{-}$. We assign these band to
the anti-symmetric twisting vibration of the two $\alpha $-particles on the $z$-axis ($\nu_6$ in Fig.~\ref{modes}). The rotational and vibrational parameters for the $K^{P}=1^{+}$ band are $B=124$ keV and $\omega=9.68$ MeV. Several members of this band appear to have been observed, including the unnatural parity states $1^{+}$ and $3^{+}$.

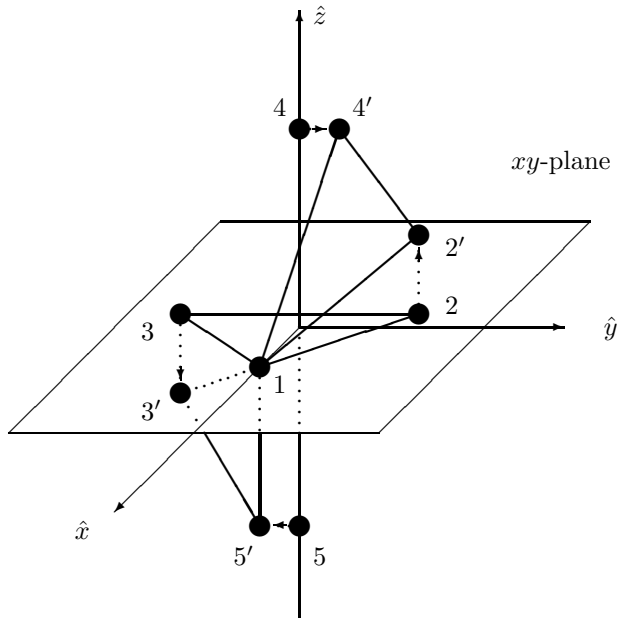
\begin{figure}
\centering
\vspace{15pt}
\setlength{\unitlength}{1pt}
\begin{picture}(360,325)(0,0)
\thinlines
\put(115,170) {\circle*{8}} 
\put(175,190) {\circle*{8}}
\put( 85,190) {\circle*{8}}
\put(130,260) {\circle*{8}}
\put(130,110) {\circle*{8}}
\put(175,220) {\circle*{8}}
\put( 85,160) {\circle*{8}}
\put(145,260) {\circle*{8}}
\put(115,110) {\circle*{8}}
\put(120,160) {$1$}
\put(185,190) {$2$}
\put( 70,180) {$3$}
\put(120,265) {$4$}
\put(135, 95) {$5$}
\put(185,212) {$2'$}
\put( 70,150) {$3'$}
\put(150,265) {$4'$}
\put(105, 95) {$5'$}
\multiput(175,190)( 0, 4){8}{\circle*{1}}
\multiput( 85,190)( 0,-4){8}{\circle*{1}}
\multiput(130,260)( 4, 0){4}{\circle*{1}}
\multiput(130,110)(-4, 0){4}{\circle*{1}}
\put(175,210) {\vector( 0, 1){5}}
\put( 85,170) {\vector( 0,-1){5}}
\put(135,260) {\vector( 1, 0){5}}
\put(125,110) {\vector(-1, 0){5}}

\put(130,185) {\vector(-1,-1){70}}
\put(130,185) {\vector( 0, 1){120}}
\put(130,185) {\vector( 1, 0){100}}
\put(130,145) {\line( 0,-1){ 70}}
\multiput(130,185)(0,-4){11}{\circle*{1}}
\put(135,300) {$\hat{z}$}
\put(245,182) {$\hat{y}$}
\put( 45,105) {$\hat{x}$}

\put(210,245) {$xy$-plane}
\put( 20,145) {\line( 1, 0){140}}
\put( 20,145) {\line( 1, 1){ 80}}
\put(160,145) {\line( 1, 1){ 80}}
\put(100,225) {\line( 1, 0){140}}

\thicklines
\put( 85,190) {\line( 1, 0){90}}
\put( 85,190) {\line( 3,-2){30}}
\put(115,170) {\line( 3, 1){60}}
\put(115,170) {\line( 6, 5){60}}
\multiput(115,170)(-3,-1){10}{\circle*{1}}
\put(115,170) {\line( 1, 3){30}}
\put(115,145) {\line( 0,-1){35}}
\multiput(115,170)( 0,-4){7}{\circle*{1}}
\put(175,220) {\line(-3, 4){30}}
\multiput(85,160)( 3,-5){3}{\circle*{1}}
\put( 94,145) {\line( 3,-5){21}}

\end{picture}
\caption{Transformation from the configuration (12345) with bipyramidal 
${\cal D}_{3h}$ symmetry towards a configuration (12'3'4'5') with distorted tetrahedral ${\cal D}_{2d}$ symmetry.}
\label{twist}
\end{figure}

The $K^{P}=2^{-}$ band has rotational and vibrational parameters $B=145$ keV and $\omega=4.10$ MeV. This band has a very low value of the vibrational parameter. Because of its unusual behavior we have investigated this vibrational mode in further detail. The geometry of the anti-symmetric twisting mode goes in the direction of the geometric structure discussed in great detail in Ref.~\cite{hauge}, that of a distorted body-centered tetrahedron with symmetry $D_{2d}$ as shown in Fig.~\ref{twist}. 
Our interpretation is supported by the large $E3$ matrix elements between members of this band and members of the ground state band, to be discussed in Section~\ref{em}. Another interpretation of this band was given by Von Oertzen \cite{vonoertzen} as a
non-collective, non-cluster, particle-hole band. However, in this interpretation is difficult to understand the large $E3$ matrix elements to the ground state band (see Section~\ref{em}).

\subsection{Summary of assignments into bands}

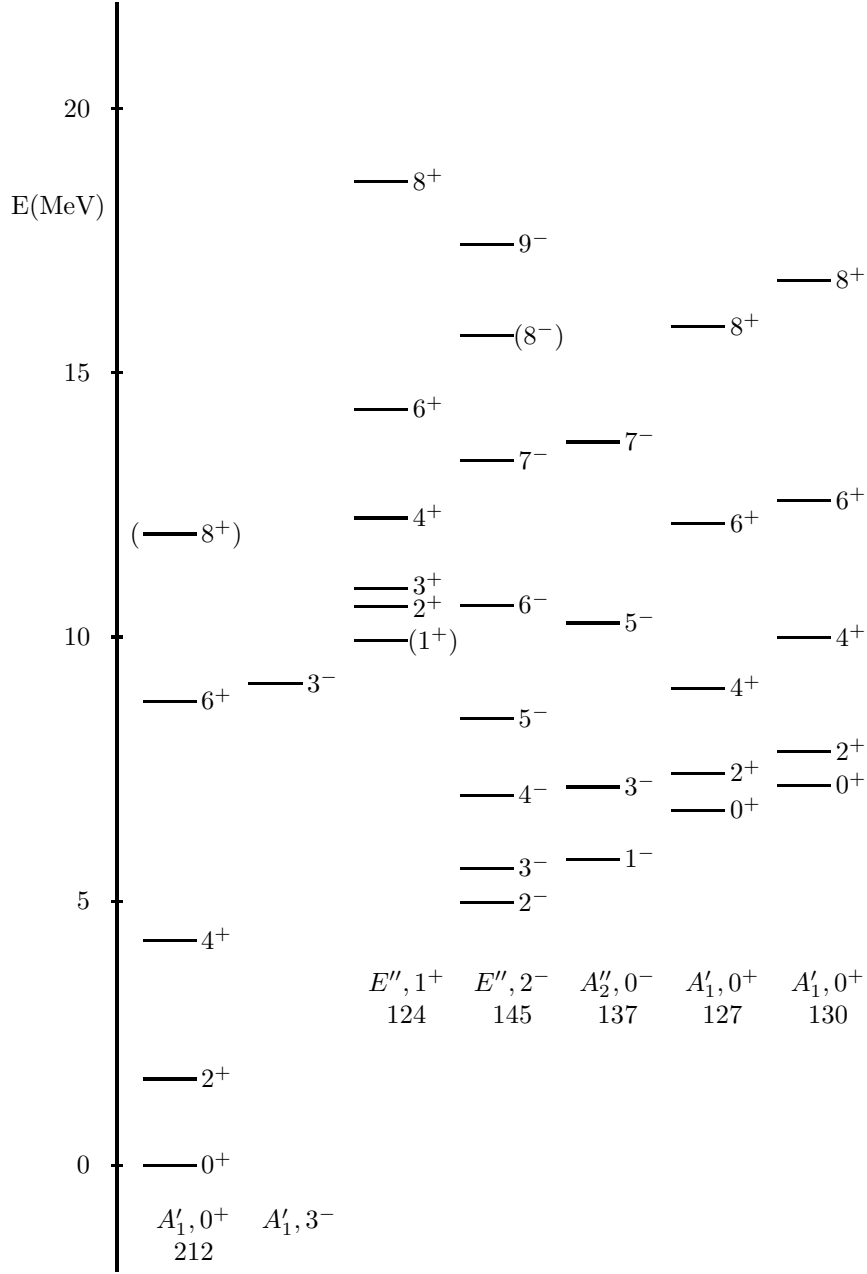
\begin{figure}
\centering
\setlength{\unitlength}{1pt}
\begin{picture}(350,480)(-10,-10)
\thicklines
\put ( 30,-10) {\line(0,1){480}}
\put ( 28, 30) {\line(1,0){4}}
\put ( 28,130) {\line(1,0){4}}
\put ( 28,230) {\line(1,0){4}}
\put ( 28,330) {\line(1,0){4}}
\put ( 28,430) {\line(1,0){4}}
\put ( 15, 27) {0}
\put ( 15,127) {5}
\put ( 10,227) {10}
\put ( 10,327) {15}
\put ( 10,427) {20}
\put (-10,390) {E(MeV)}
\put( 40, 30.0) {\line(1,0){20}}
\put( 40, 62.6) {\line(1,0){20}}
\put( 40,115.0) {\line(1,0){20}}
\put( 40,205.6) {\line(1,0){20}}
\put( 40,269.0) {\line(1,0){20}}
\put( 40,  0.0) {$\begin{array}{c} A'_1, 0^+ \\ 212 \end{array}$}
\put( 62, 27.0) {$0^+$}
\put( 62, 59.6) {$2^+$}
\put( 62,112.0) {$4^+$}
\put( 62,202.6) {$6^+$}
\put( 35,266.0) {$($}
\put( 62,266.0) {$8^+)$}
\put( 80,212.2) {\line(1,0){20}}
\put( 80,  0.0) {$\begin{array}{c} A'_1, 3^- \\ \mbox{} \end{array}$}
\put(102,209.2) {$3^-$}
\put(120,228.6) {\line(1,0){20}}
\put(120,241.6) {\line(1,0){20}}
\put(120,248.4) {\line(1,0){20}}
\put(120,275.0) {\line(1,0){20}}
\put(120,316.2) {\line(1,0){20}}
\put(120,402.4) {\line(1,0){20}}
\put(120, 90.0) {$\begin{array}{c} E'', 1^+ \\ 124 \end{array}$}
\put(140,225.6) {$(1^+)$}
\put(142,237.6) {$2^+$}
\put(142,246.4) {$3^+$}
\put(142,272.0) {$4^+$}
\put(142,313.2) {$6^+$}
\put(142,399.4) {$8^+$}
\put(160,129.4) {\line(1,0){20}}
\put(160,142.4) {\line(1,0){20}}
\put(160,170.0) {\line(1,0){20}}
\put(160,199.0) {\line(1,0){20}}
\put(160,242.0) {\line(1,0){20}}
\put(160,296.8) {\line(1,0){20}}
\put(160,344.0) {\line(1,0){20}}
\put(160,378.6) {\line(1,0){20}}
\put(160, 90.0) {$\begin{array}{c} E'', 2^- \\ 145 \end{array}$}
\put(182,126.4) {$2^-$}
\put(182,139.4) {$3^-$}
\put(182,167.0) {$4^-$}
\put(182,196.0) {$5^-$}
\put(182,239.0) {$6^-$}
\put(182,293.8) {$7^-$}
\put(180,341.0) {$(8^-)$}
\put(182,375.6) {$9^-$}
\put(200,145.8) {\line(1,0){20}}
\put(200,173.2) {\line(1,0){20}}
\put(200,235.2) {\line(1,0){20}}
\put(200,303.8) {\line(1,0){20}}
\put(200, 90.0) {$\begin{array}{c} A''_2, 0^- \\ 137 \end{array}$}
\put(222,142.8) {$1^-$}
\put(222,170.2) {$3^-$}
\put(222,232.2) {$5^-$}
\put(222,300.8) {$7^-$}
\put(240,164.4) {\line(1,0){20}}
\put(240,178.4) {\line(1,0){20}}
\put(240,210.6) {\line(1,0){20}}
\put(240,272.8) {\line(1,0){20}}
\put(240,347.4) {\line(1,0){20}}
\put(240, 90.0) {$\begin{array}{c} A'_1, 0^+ \\ 127 \end{array}$}
\put(262,161.4) {$0^+$}
\put(262,175.4) {$2^+$}
\put(262,207.6) {$4^+$}
\put(262,269.8) {$6^+$}
\put(262,344.4) {$8^+$}
\put(280,173.8) {\line(1,0){20}}
\put(280,186.6) {\line(1,0){20}}
\put(280,229.8) {\line(1,0){20}}
\put(280,281.6) {\line(1,0){20}}
\put(280,365.0) {\line(1,0){20}}
\put(280, 90.0) {$\begin{array}{c} A'_1, 0^+ \\ 130 \end{array}$}
\put(302,170.8) {$0^+$}
\put(302,183.6) {$2^+$}
\put(302,226.8) {$4^+$}
\put(302,278.6) {$6^+$}
\put(302,362.0) {$8^+$}
\end{picture}
\caption{Rotational bands of the ground state and the nine vibrations in $^{20}$Ne. The bands are labeled by $\Gamma$, $K^P$ and the value of the rotational parameter $B$.}
\label{bands}
\end{figure}

\addtocounter{figure}{-1}
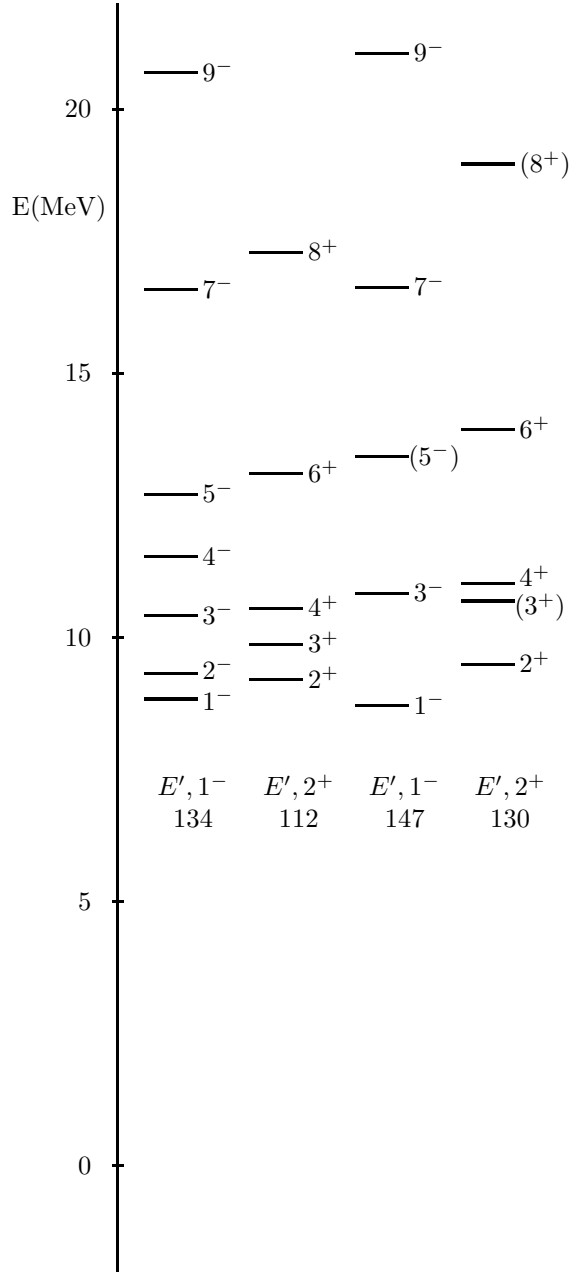
\begin{figure}
\centering
\setlength{\unitlength}{1pt}
\begin{picture}(230,480)(-10,-10)
\thicklines
\put ( 30,-10) {\line(0,1){480}}
\put ( 28, 30) {\line(1,0){4}}
\put ( 28,130) {\line(1,0){4}}
\put ( 28,230) {\line(1,0){4}}
\put ( 28,330) {\line(1,0){4}}
\put ( 28,430) {\line(1,0){4}}
\put ( 15, 27) {0}
\put ( 15,127) {5}
\put ( 10,227) {10}
\put ( 10,327) {15}
\put ( 10,427) {20}
\put (-10,390) {E(MeV)}
\put( 40,206.8) {\line(1,0){20}}
\put( 40,216.4) {\line(1,0){20}}
\put( 40,238.2) {\line(1,0){20}}
\put( 40,260.6) {\line(1,0){20}}
\put( 40,284.2) {\line(1,0){20}}
\put( 40,361.6) {\line(1,0){20}}
\put( 40,443.8) {\line(1,0){20}}
\put( 40,164.2) {$\begin{array}{c} E', 1^- \\ 134 \end{array}$}
\put( 62,202.8) {$1^-$}
\put( 62,214.4) {$2^-$}
\put( 62,235.2) {$3^-$}
\put( 62,257.6) {$4^-$}
\put( 62,281.2) {$5^-$}
\put( 62,358.6) {$7^-$}
\put( 62,440.8) {$9^-$}
\put( 80,214.0) {\line(1,0){20}}
\put( 80,227.4) {\line(1,0){20}}
\put( 80,241.0) {\line(1,0){20}}
\put( 80,292.0) {\line(1,0){20}}
\put( 80,375.8) {\line(1,0){20}}
\put( 80,164.2) {$\begin{array}{c} E', 2^+ \\ 112 \end{array}$}
\put(102,211.0) {$2^+$}
\put(102,224.4) {$3^+$}
\put(102,238.0) {$4^+$}
\put(102,289.0) {$6^+$}
\put(102,372.8) {$8^+$}
\put(120,204.2) {\line(1,0){20}}
\put(120,246.8) {\line(1,0){20}}
\put(120,298.4) {\line(1,0){20}}
\put(120,362.6) {\line(1,0){20}}
\put(120,451.2) {\line(1,0){20}}
\put(120,164.2) {$\begin{array}{c} E', 1^- \\ 147 \end{array}$}
\put(142,201.2) {$1^-$}
\put(142,243.8) {$3^-$}
\put(140,295.4) {$(5^-)$}
\put(142,359.6) {$7^-$}
\put(142,448.2) {$9^-$}
\put(160,219.8) {\line(1,0){20}}
\put(160,243.8) {\line(1,0){20}}
\put(160,250.4) {\line(1,0){20}}
\put(160,308.6) {\line(1,0){20}}
\put(160,409.2) {\line(1,0){20}}
\put(160,164.2) {$\begin{array}{c} E', 2^+ \\ 130 \end{array}$}
\put(182,216.8) {$2^+$}
\put(180,238.8) {$(3^+)$}
\put(182,249.4) {$4^+$}
\put(182,305.6) {$6^+$}
\put(182,406.2) {$(8^+)$}
\end{picture}
\caption{Continued.}
\end{figure}

In addition to the ground state band, we have found evidence for all nine expected vibrational bands of the bi-pyramidal configuration. There are three one-dimensional modes corresponding to the anti-symmetric stretching of the two $\alpha$-particles on the $z$-axis with $A''_2$ symmetry ($K^{P}=0^{-}(5.79)$ band), the symmetric stretching of the two $\alpha$-particles on the $z$-axis with $A'_1$ symmetry 
($K^{P}=0^{+}(6.72)$ band), and the stretching vibration of the triangle with $A_1'$ symmetry ($K^{P}=0^{+}(7.19)$ band). In addition, there are three two-dimensional modes corresponding to the two-dimensional bending vibration of the triangle with $E'$ symmetry ($K^{P}=1^{-}(8.84)$ and $K^{P}=2^{+}(9.20)$ band), the two-dimensional bending vibration of the two $\alpha$-particles on the $z$-axis with $E'$ symmetry ($K^{P}=1^{-}(8.71)$ and $K^{P}=2^{+}(9.49)$ bands), and the two-dimensional twisting vibration with $E''$ symmetry ($K^{P}=1^{+}(9.93)$ and $K^{P}=2^{-}(4.97)$ bands). 
It is noteworthy that all expected vibrational bands and no other have been
observed with band-head energies below $10$ MeV. The only unusual behavior is
that of the band $K^{P}=2^{-}(4.97)$ which is at very low energy. The
results are summarized in Table~\ref{gsvib} and Fig.~\ref{bands}.

\begin{table}
\centering
\caption{Summary of assignments of rotational bands.}
\label{gsvib}
\vspace{10pt}
\begin{tabular}{ccccc}
\hline
\noalign{\smallskip}
& $\Gamma$ & $K^P$ & $\omega$ (MeV) & $B$ (keV) \\
\noalign{\smallskip}
\hline
\noalign{\smallskip}
g.s.    & $A'_1$  & $0^+$ & 0.00 & 212 \\ 
$v_{1}$ & $A''_2$ & $0^-$ & 5.52 & 137 \\
$v_{2}$ & $A'_1$  & $0^+$ & 6.72 & 127 \\
$v_{3}$ & $A'_1$  & $0^+$ & 7.19 & 130 \\ 
$v_{4}$ & $E'$    & $1^-$ & 8.59 & 134 \\
        &         & $2^+$ & 8.53 & 112 \\
$v_{5}$ & $E'$    & $1^-$ & 8.42 & 147 \\ 
        &         & $2^+$ & 8.71 & 130 \\ 
$v_{6}$ & $E''$   & $1^+$ & 9.68 & 124 \\
        &         & $2^-$ & 4.10 & 145 \\ 
\noalign{\smallskip}
\hline
\end{tabular}
\end{table}

The rotational constants, $B$, and vibrational energies, $\omega$, are
remarkable in many ways. The rotational constants of the vibrational modes
are all comparable to each other, $B \cong 130$ keV, and about 60\% of the
rotational constant of the ground state band, $B=212$ keV. This is in line
with other cluster studies, as those in $^{12}$C \cite{bijker2} and in $%
^{16}$O \cite{bijker4}. The geometry of the vibrations is more extended
that that of the ground state with larger moment of inertia. The vibrational
energies are all comparable to each other with the exception of the $%
\omega_{6}$ mode which appears to be much softer. The occurrence of the
rotational bands $K^{P}=0^{+}(7.19)$ and $K^{P}=1^{-}(8.84)$ is strong evidence
that the triangle ($^{12}$C) is a substructure of $^{20}$Ne, and that the
bi-pyramidal structure is appropriate for this nucleus, as suggested many
years ago by Brink \cite{brink2}.

\section{Electromagnetic transition rates in $^{20}$Ne}
\label{em}

Electromagnetic transition rates of the ground state band $A_{1}^{\prime
}:K^{P}=0^{+}(0.0)$ have been analyzed in Section~\ref{emgs}. In this section we
consider the available information \cite{tilley} for the other bands in
order of increasing excitation energy.

\subsection{$E'':K^{P}=2^{-}(4.97)$}

Some $E2$ and $E3$ transition rates have been measured for this band. We
quote the experimental values both in W.U. and in $e^{2}$fm$^{2\lambda }$ to
emphasize their collective behavior. We use the notation $B\left( E\lambda
;K^{\prime },L^{\prime }\rightarrow K,L\right)$.

\subsubsection{$B(E2)$,	$\Delta K=0$}

The available data are given in Table~\ref{be8} where they are compared with the
values calculated using
\begin{equation}
B(E2;K',L' \rightarrow K,L) \;=\; \frac{5}{16\pi} \, Q_0^2 \, 
\left\langle L^{\prime},K^{\prime},2,0|L,K\right\rangle^{2} ~,
\label{be2}
\end{equation}%
and the value of $Q_{0}$ from the ground state $Q_{0,g.s.}=52.5$ efm$^{2}$.
We see that the calculated values are in agreement with the data for the
transitions $5^{-}\rightarrow 3^{-}$ and $6^{-}\rightarrow 4^{-}$ but in
disagreement for $4^{-}\rightarrow 2^{-}$. The latter value is quoted in 
\cite{tilley} without error and as uncertain.

\begin{table}
\centering
\caption{$B(E2;K^{\prime},L^{\prime}\rightarrow K,L)$ values.}
\vspace{10pt}
\label{be8}
\begin{tabular}{lccc}
\hline
\noalign{\smallskip}
& \multicolumn{2}{c}{Exp} & Calc \\
& W.U. & e$^2$fm$^4$ & e$^2$fm$^4$ \\ 
\noalign{\smallskip}
\hline
\noalign{\smallskip}
$B(E2;2^{-}(4.97),4^{-} \rightarrow 2^{-}(4.97),2^{-})$ & 1.8  & 5.8 & 33 \\ 
$B(E2;2^{-}(4.97),5^{-} \rightarrow 2^{-}(4.97),3^{-})$ & 27(6) & 86.9(19.3) & 52 \\ 
$B(E2;2^{-}(4.97),6^{-} \rightarrow 2^{-}(4.97),4^{-})$ & 17(6) & 54.7(19.3) & 64 \\
\noalign{\smallskip}
\hline
\end{tabular}
\end{table}

\subsubsection{$B(E3),$ $\Delta K=2$}

The available data are given in Table 9. These transitions, being $\Delta
K=2 $ are calculated using%
\begin{equation}
B(E3;K^{\prime },L^{\prime }\rightarrow K,L)=2\left( Q_{32}\right)
^{2}\left\langle L^{\prime },K^{\prime },3,2|L,K\right\rangle ^{2}.
\end{equation}%
Here $Q_{32}$ is the transition moment from $K^{P}=2^{-}(4.97)$ to $%
K^{P}=0^{+}(0.0)$. Using a value of $Q_{32}=16.3$ $e$fm$^{3}$ we obtain the
calculated values of Table~\ref{be9}. These values are inconsistent with each other
by a factor of $\sim 2$.

\begin{table}
\centering
\caption{$B(E3;K^{\prime},L^{\prime}\rightarrow K,L)$ values.}
\vspace{10pt}
\label{be9}
\begin{tabular}{lccc}
\hline
\noalign{\smallskip}
& \multicolumn{2}{c}{Exp} & Calc \\
& W.U. & e$^2$fm$^6$ & e$^2$fm$^6$ \\ 
\noalign{\smallskip}
\hline
\noalign{\smallskip}
$B(E3;2^{-}(4.97),2^{-} \rightarrow 0^{+}(0.0),2^{+})$ & 6(2)  & 143(47) & 190 \\ 
$B(E3;2^{-}(4.97),3^{-} \rightarrow 0^{+}(0.0),0^{+})$ & 11(4) & 261(95) &  76 \\
\noalign{\smallskip}
\hline
\end{tabular}
\end{table}

The value of $B(E3;0^{+},0^{+}\rightarrow 2^{-},3^{-})$ can also be
estimated if we assume that the $K^{P}=2^{-}(4.97)$ band is the vibration of
Fig.~\ref{twist}. A complicated derivation \cite{bijker-tobepublished} gives%
\begin{equation}
B(E3;0^{+},0^{+}\rightarrow 2^{-},3^{-}) \;=\; \frac{105}{32\pi}(2e)^{2}\beta
_{t}^{6}\frac{3}{2} ~,
\end{equation}%
where $\beta_{t}$ is the transition radius. By using the value of $Q_{32}=16.3$ $e$fm$^{3}$, we then obtain $\beta_{t}=2.22$ fm, which is in
line with the values of $\beta_{1}$ and $\beta_{2}$ of the ground state
configuration. In any event, the large $B(E3)$ values of Table~\ref{be9} are not
consistent with a particle-hole interpretation of the $K^{P}=2^{-}(4.97)$ band, but instead to a collective vibrational interpretation of this band.

\subsection{$A_{2}^{\prime \prime }:K^{P}=0^{-}(5.79)$}

\subsubsection{$B(E2)$, $\Delta K=0$}

Only one transition is known as shown in Table~\ref{be10}. One can calculate this
transition using Eq.~(\ref{be2}) and the value of $Q_{0}$ from the ground state $Q_{0,g.s.}=52.5$ efm$^{2}$, obtaining the value given in Table~\ref{be10}.

\begin{table}
\centering
\caption{$B(E2;K^{\prime},L^{\prime}\rightarrow K,L)$ values.}
\vspace{10pt}
\label{be10}
\begin{tabular}{lccc}
\hline
\noalign{\smallskip}
& \multicolumn{2}{c}{Exp} & Calc \\
& W.U. & e$^2$fm$^4$ & e$^2$fm$^4$ \\ 
\noalign{\smallskip}
\hline
\noalign{\smallskip}
$B(E2;0^{-}(5.79),3^{-} \rightarrow 0^{-}(5.79),1^{-})$ & 50(8) & 161(26) & 71 \\
\noalign{\smallskip}
\hline
\end{tabular}
\end{table}

It appears that this band is even more collective than the ground state
band, as expected from our interpretation as the anti-symmetric stretching
vibration of the two $\alpha $-particles on the $z$-axis. (The two particles
lie at a greater distance from the center than in the ground state.) Its
value of $Q_{0}$ appears to be 79.3 efm$^{2}$.

\subsection{$A_{1}^{\prime }:K^{P}=0^+(6.72)$}

For this band only out-of-band transitions have been measured. These
transitions can be analyzed with the formula%
\begin{equation}
B\left( E2;0^{+},L^{\prime }\rightarrow 0^{+},L\right) \;=\; 
\frac{5}{16\pi} \, Q_{0,t}^{2} \, 
\left\langle L^{\prime },0,2,0|L,0\right\rangle^{2} ~,
\label{be2t}
\end{equation}%
where $Q_{0,t}$ is the transition moment. This calculation is compared in
Table~\ref{be11} withe experiment. The extracted value of $Q_{0t}$ is $Q_{0t}=10.7$ 
efm$^{2}$.

\begin{table}
\centering
\caption{$B(E2;K^{\prime},L^{\prime}\rightarrow K,L)$ values.}
\vspace{10pt}
\label{be11}
\begin{tabular}{lccc}
\hline
\noalign{\smallskip}
& \multicolumn{2}{c}{Exp} & Calc \\
& W.U. & e$^2$fm$^4$ & e$^2$fm$^4$ \\ 
\noalign{\smallskip}
\hline
\noalign{\smallskip}
$B(E2;0^{+}(6.72),0^{+} \rightarrow 0^{+}(0.0),2^{+})$ & 3.6?   & 11.5    & 11.5 \\ 
$B(E2;0^{+}(6.72),2^{+} \rightarrow 0^{+}(0.0),2^{+})$ & 1.7(2) &  5.5(6) &  5.9 \\ 
$B(E2;0^{+}(6.72),4^{+} \rightarrow 0^{+}(0.0),2^{+})$ & 5.8(7) & 18.7(22)&  3.3 \\
\noalign{\smallskip}
\hline
\end{tabular}
\end{table}

\subsection{$A_{1}^{\prime }:K^{P}=0^{+}(7.19)$}

For this band only out-of-band transitions have been measured. In Table~\ref{be12},
their calculation, using Eq.~(\ref{be2t}) and $Q_{0t}=10.7$ efm$^{2}$ as before, is
given. While the decays of $K^{P}=0^{+}(6.72)$ and $K^{P}=0^{+}(7.19)$ are
consistent with each other in having the same transition moment, $Q_{ot}$,
they show substantial differences with experiment, especially the
transitions from the $4^{+}$ states which appear to be more collective than
the calculation shows.

\begin{table}
\centering
\caption{$B(E2;K^{\prime},L^{\prime}\rightarrow K,L)$ values.}
\vspace{10pt}
\label{be12}
\begin{tabular}{lccc}
\hline
\noalign{\smallskip}
& \multicolumn{2}{c}{Exp} & Calc \\
& W.U. & e$^2$fm$^4$ & e$^2$fm$^4$ \\ 
\noalign{\smallskip}
\hline
\noalign{\smallskip}
$B(E2;0^{+}(7.19),0^{+} \rightarrow 0^{+}(0.0),2^{+})$ & 0.31(6) & 1.0(2) & 11.5 \\ 
$B(E2;0^{+}(7.19),2^{+} \rightarrow 0^{+}(0.0),0^{+})$ & 0.73(9) & 2.3(3) &  2.3 \\ 
$B(E2;0^{+}(7.19),4^{+} \rightarrow 0^{+}(0.0),2^{+})$ & 8.3(37) & 26.7(119) & 3.3 \\
\noalign{\smallskip}
\hline
\end{tabular}
\end{table}

\subsection{Summary of electromagnetic transitions}

The in-band transitions show large $E2$ collectivity with intrinsic
quadrupole moments $Q_{0}$ as in Table~\ref{be13}. The out-of-band transitions show
small collectivity with transition moments $Q_{0t}$ as in Table~\ref{be13}.

\begin{table}
\centering
\caption{Summary of $E2$ transitions.}
\vspace{10pt}
\label{be13}
\begin{tabular}{cc}
\hline
\noalign{\smallskip}
Intraband & $Q_0$ (efm$^2$) \\ 
\noalign{\smallskip}
\hline
\noalign{\smallskip}
g.s.($A'_{1}$)         & 52.5 \\
$\omega_{2}(A''_{2}$)  & 79.3 \\
$\omega_{6}(E''$)      & 52.5 \\ 
\noalign{\smallskip}
\hline
\noalign{\smallskip}
Interband & $Q_{0,{\rm t}}$ (efm$^2$) \\ 
\noalign{\smallskip}
\hline
\noalign{\smallskip}
$\omega_{2}(A'_{1}) \rightarrow g.s.(A'_{1})$ & 10.7 \\
$\omega_{3}(A'_{1}) \rightarrow g.s.(A'_{1})$ & 10.7 \\
\noalign{\smallskip}
\hline
\end{tabular}
\end{table}

\section{Double excitation spectrum of $^{20}$Ne: Further evidence for ${\cal D}_{3h}$ symmetry}

In a given molecular structure one expects double vibrational modes. These
modes can be obtained by multiplication of the irreducible representations
of the corresponding discrete group. The irreducible representations of ${\cal D}_{3h}$ were given in Table~\ref{gamma}. In Table~\ref{multiplication} we give the multiplication table of ${\cal D}_{3h}$ in the notation of \cite{herzberg}.

\begin{table}
\centering
\caption{Multiplication table of ${\cal D}_{3h}$.}
\vspace{10pt}
\label{multiplication}
\begin{tabular}{c|cccccc}
\hline
\noalign{\smallskip}
& $A'_{1}$ & $A'_{2}$ & $A''_{1}$ & $A''_{2}$ & $E'$ & $E''$ \\ 
\noalign{\smallskip}
\hline
\noalign{\smallskip}
$A'_{1}$  & $A'_{1}$ & $A'_{2}$ & $A''_{1}$ & $A''_{2}$ & $E'$  & $E''$ \\ 
$A'_{2}$  &          & $A'_{1}$ & $A''_{2}$ & $A''_{1}$ & $E'$  & $E''$ \\ 
$A''_{1}$ &          &          & $A'_{1}$  & $A'_{2}$  & $E''$ & $E'$  \\ 
$A''_{2}$ &          &          &           & $A'_{1}$  & $E''$ & $E'$  \\
$E'$      &          &          &           &           
& $A'_{1}$, $A'_{2}$, $E'$ & $A''_{1}$, $A''_{2}$, $E''$ \\ 
$E''$     &          &          &           &           &       
& $A'_{1}$, $A'_{2}$, $E'$ \\ 
\noalign{\smallskip}
\hline
\end{tabular}
\end{table}

The energy levels of a bi-pyramidal configuration with ${\cal D}_{3h}$ symmetry,
including double excitations, can now be written as%
\begin{eqnarray}
E\left( \left[ v\right] ,K,L\right) &=& E_{0} 
+ \sum_{i=1}^{6} \left( \omega _{i}-x_{ii} \right) v_{i} 
+ \sum_{i<j=1}^{6} x_{ij} v_{i} v_{j}
\nonumber\\
&& + B_{x\left[v\right]} L(L+1) 
+ \left(B_{z}-B_{x}\right)_{\left[v\right]} K^{2} 
\end{eqnarray}
where we have added to Eq.~(\ref{energies}) the anharmonic terms with anharmonicity
constants $x_{ij}$.

\subsection{Assignments of states to doubly excited bands and their ${\cal D}_{3h}$
classification}

Assignments to bands in addition to those of Section~\ref{spectrum} have been done by Tilley \textit{et al.} \cite{tilley} and by several other authors. We report in
Table~\ref{doublevib} those bands for which assignments can be done with some confidence.

\begin{table}
\centering
\caption{Assignments of bands to double vibrations.}
\label{doublevib}
\vspace{10pt}
\begin{tabular}{ccccrrr}
\hline
\noalign{\smallskip}
$[v_{1},v_{2},v_{3},v_{4},v_{5},v_{6}]$ & $\Gamma$ & $K^{P}$ & $L^{P}$ 
& $E_{\rm exp}$ & $E_{\rm th}$ & $E_{\rm exp}$ \cite{tilley} \\ 
\noalign{\smallskip}
\hline
\noalign{\smallskip}
$[2,0,0,0,0,0]$ & $A'_{1}$ & $0^{+}(8.7)$ 
    & $0^{+}$ &   8.7   &  8.7 &   8.7   \\ 
& & & $2^{+}$ &   9.0   &  9.3 &   8.8   \\ 
& & & $4^{+}$ &  10.8   & 10.8 &  10.8   \\ 
& & & $6^{+}$ & (14.81) & 13.1 & (12.59) \\ 
& & & $8^{+}$ &         & 16.3 & (17.30) \\ 
\noalign{\smallskip}
$[0,2,0,0,0,0]$ & $A'_{1}$ & $0^{+}(10.97)$ 
    & $0^+$ & 10.97 & 11.0 & 10.97 \\ 
& & & $2^+$ & 12.33 & 11.7 & 12.33 \\ 
& & & $4^+$ & 13.34 & 13.3 &       \\ 
& & & $6^+$ & 15.16 & 15.7 &       \\ 
& & & $8^+$ & 19.73 & 19.3 &       \\ 
\noalign{\smallskip}
$[0,1,1,0,0,0]$ & $A'_{1}$ & $0^{+}(11.55)$  
    & $0^+$ &  11.55  & 11.5 &  11.55  \\ 
& & & $2^+$ &  11.88  & 12.3 &         \\ 
& & & $4^+$ &  13.97  & 14.0 & (13.97) \\ 
& & & $6^+$ &  16.51  & 16.6 & (16.51) \\ 
& & & $8^+$ & (18.54) & 20.3 & (18.62) \\ 
\noalign{\smallskip}
$[0,0,2,0,0,0]$ & $A'_{1}$ & $0^{+}(12.43)$  
    & $0^+$ & 12.43 & 12.4 &  12.43  \\ 
& & & $2^+$ & 12.96 & 13.3 & (12.96) \\ 
& & & $4^+$ & 15.33 & 15.3 &         \\ 
& & & $6^+$ & 19.44 & 18.5 & (19.44) \\ 
& & & $8^+$ &       & 22.9 &         \\ 
\noalign{\smallskip}
$[1,1,0,0,0,0]$ & $A''_{2}$ & $0^{-}(11.24)$ 
    & $1^-$ & 11.24 & 11.2 &  \\ 
& & & $3^-$ & 12.25 & 12.2 &  \\ 
& & & $5^-$ & 13.68 & 14.1 &  \\ 
& & & $7^-$ & 15.37 & 15.3 &  \\ 
& & & $9^-$ &       & 20.1 &  \\ 
\noalign{\smallskip}
$[1,0,1,0,0,0]$ & $A''_{2}$ & $0^{-}(11.27)$ 
    & $1^-$ & 11.27 & 11.3 &  \\ 
& & & $3^-$ & 12.40 & 12.4 &  \\ 
& & & $5^-$ & 14.45 & 14.4 &  \\ 
& & & $7^-$ & 18.00 & 15.8 &  \\ 
\noalign{\smallskip}
\hline
\end{tabular}
\end{table}

\subsubsection{$A_{2}^{\prime \prime }(5.79)\times A_{2}^{\prime \prime
}(5.79)=A_{1}^{\prime }:K^{P}=0^{+}(8.7)$}

This band was assigned in \cite{tilley}. Our assignments slightly differ
from those of \cite{tilley}. The rotational and anharmonicity parameters are 
$B=105$ keV, $x=-0.78$ MeV. The vibrational parameter $\omega =5.52$ MeV is
taken from Table~\ref{gsvib}. This band is interpreted as the double
anti-symmetric stretching vibration of the two $\alpha$-particles on the $z$-axis.

\subsubsection{$A_{1}^{\prime }(6.72)\times A_{1}^{\prime
}(6.72)=A_{1}^{\prime}:K^{P}=0^{+}(10.97)$}

The rotational and anharmonicity parameters of this band are $B=118$ keV, $%
x=-0.82$ MeV. The vibrational parameter from Table~\ref{gsvib} is $\omega =6.72$
MeV. This band is interpreted as the double symmetric stretching vibration.

\subsubsection{$A_{1}^{\prime }(6.72)\times A_{1}^{\prime
}(7.19)=A_{1}^{\prime }:K^{P}=0^{+}(11.55)$}

The rotational and anharmonicity parameters are $B=121$ keV, $x=-0.89$ MeV.
The vibrational parameters are taken from Table~\ref{gsvib}.

\subsubsection{$A_{1}^{\prime }(7.19)\times A_{1}^{\prime
}(7.19)=A_{1}^{\prime }:K^{P}=0^{+}(12.43)$}

The rotational and anharmonicity parameters are $B=145$ keV, $x=-0.65$. The
vibrational parameters are taken from Table~\ref{gsvib}. This band is interpreted as
the double stretching vibration of the triangle.

\subsubsection{$A_{2}^{\prime \prime }(5.79)\times A_{1}^{\prime
}(6.72)=A_{2}^{\prime \prime }:K^{P}=0^{-}(11.24)$}

This band was not assigned in \cite{tilley}. The rotational and
anharmonicity parameters are $B=101$ keV, $x=+0.60$ MeV. Note the positive
sign, typical of combination bands. The vibrational parameters are taken from Table~\ref{gsvib}.

\subsubsection{$A_{2}^{\prime \prime }(5.79)\times A_{1}^{\prime
}(7.19)=A_{2}^{\prime \prime }:K^{P}=0^{-}(11.27)$}

This band was also not assigned in \cite{tilley}. The rotational and
anharmonicity parameters are $B=113$ keV and $x=-0.01$ MeV.

\subsection{Summary of assignments of double bands}

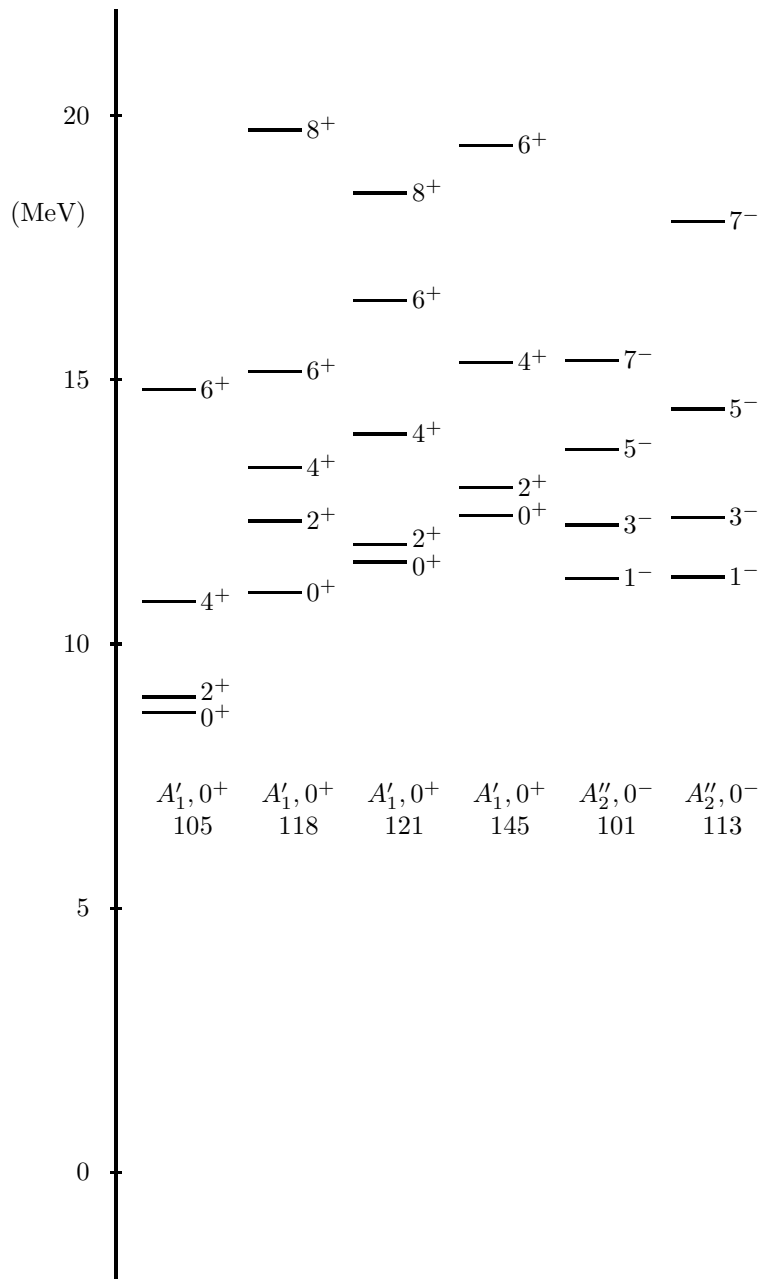
\begin{figure}
\centering
\setlength{\unitlength}{1pt}
\begin{picture}(310,480)(-10,-10)
\thicklines
\put ( 30,-10) {\line(0,1){480}}
\put ( 28, 30) {\line(1,0){4}}
\put ( 28,130) {\line(1,0){4}}
\put ( 28,230) {\line(1,0){4}}
\put ( 28,330) {\line(1,0){4}}
\put ( 28,430) {\line(1,0){4}}
\put ( 15, 27) {0}
\put ( 15,127) {5}
\put ( 10,227) {10}
\put ( 10,327) {15}
\put ( 10,427) {20}
\put (-10,390) {(MeV)}
\put( 40,204.0) {\line(1,0){20}}
\put( 40,210.0) {\line(1,0){20}}
\put( 40,246.0) {\line(1,0){20}}
\put( 40,326.2) {\line(1,0){20}}
\put( 40,164.0) {$\begin{array}{c} A'_1, 0^+ \\ 105 \end{array}$}
\put( 62,199.0) {$0^+$}
\put( 62,209.0) {$2^+$}
\put( 62,243.0) {$4^+$}
\put( 62,323.2) {$6^+$}
\put( 80,249.4) {\line(1,0){20}}
\put( 80,276.6) {\line(1,0){20}}
\put( 80,296.8) {\line(1,0){20}}
\put( 80,333.2) {\line(1,0){20}}
\put( 80,424.6) {\line(1,0){20}}
\put( 80,164.0) {$\begin{array}{c} A'_1, 0^+ \\ 118 \end{array}$}
\put(102,246.4) {$0^+$}
\put(102,273.6) {$2^+$}
\put(102,293.8) {$4^+$}
\put(102,330.2) {$6^+$}
\put(102,421.6) {$8^+$}
\put(120,261.0) {\line(1,0){20}}
\put(120,267.6) {\line(1,0){20}}
\put(120,309.4) {\line(1,0){20}}
\put(120,360.2) {\line(1,0){20}}
\put(120,400.8) {\line(1,0){20}}
\put(120,164.0) {$\begin{array}{c} A'_1, 0^+ \\ 121 \end{array}$}
\put(142,256.0) {$0^+$}
\put(142,266.6) {$2^+$}
\put(142,306.4) {$4^+$}
\put(142,357.2) {$6^+$}
\put(142,397.8) {$8^+$}
\put(160,278.6) {\line(1,0){20}}
\put(160,289.2) {\line(1,0){20}}
\put(160,336.6) {\line(1,0){20}}
\put(160,418.8) {\line(1,0){20}}
\put(160,164.0) {$\begin{array}{c} A'_1, 0^+ \\ 145 \end{array}$}
\put(182,275.6) {$0^+$}
\put(182,286.2) {$2^+$}
\put(182,333.6) {$4^+$}
\put(182,415.8) {$6^+$}
\put(200,254.8) {\line(1,0){20}}
\put(200,275.0) {\line(1,0){20}}
\put(200,303.6) {\line(1,0){20}}
\put(200,337.4) {\line(1,0){20}}
\put(200,164.0) {$\begin{array}{c} A''_2, 0^- \\ 101 \end{array}$}
\put(222,251.8) {$1^-$}
\put(222,272.0) {$3^-$}
\put(222,300.6) {$5^-$}
\put(222,334.4) {$7^-$}
\put(240,255.4) {\line(1,0){20}}
\put(240,278.0) {\line(1,0){20}}
\put(240,319.0) {\line(1,0){20}}
\put(240,390.0) {\line(1,0){20}}
\put(240,164.0) {$\begin{array}{c} A''_2, 0^- \\ 113 \end{array}$}
\put(262,252.4) {$1^-$}
\put(262,275.0) {$3^-$}
\put(262,316.0) {$5^-$}
\put(262,387.0) {$7^-$}
\end{picture}
\caption{Rotational bands of the six double vibrations in $^{20}$Ne. The bands are labeled by $\Gamma$, $K^P$ and the value of the rotational parameter $B$.}
\label{vibvib}
\end{figure}

We have found evidence for six double vibrations of the bi-byramidal
configuration, $K^{P}=0^{+}(8.71)$, $K^{P}=0^{+}(10.97)$, $%
K^{P}=0^{+}(11.55) $, $K^{P}=0^{+}(12.43)$, $K^{P}=0^{-}(11.24)$, $%
K^{P}=0^{-}(11.27)$. These are all the bands expected as double vibrations
of the bands $K^{P}=0^{-}(5.79)$, $K^{P}=0^{+}(6.72)$, $K^{P}=0^{+}(7.19)$.
It is remarkable that all six expected bands have been observed. The results
are summarized in Table~\ref{twovib} and Fig.~\ref{vibvib}. The rotational parameters are almost all equal and the same as those of the vibrations in Table~\ref{bands}. The anharmonicity parameters of the positive parity bands are almost all equal and negative, while those of the negative parity combination bands are positive or small.

\begin{table}
\centering
\caption{Summary of assignments into double bands.}
\label{twovib}
\vspace{10pt}
\begin{tabular}{llcrr}
\hline
\noalign{\smallskip}
& $\Gamma$ & $E_{\rm vib}$ & $x$ (MeV) & $B$ (keV) \\
\noalign{\smallskip}
\hline
\noalign{\smallskip}
$A''_{2} \times A''_{2}$ & $A'_1$  & $2\omega_{1}$           & $-0.78$ & 105 \\
$A'_{1}  \times A'_{1}$  & $A'_1$  & $2\omega_{2}$           & $-0.82$ & 118 \\
$A'_{1}  \times A'_{1}$  & $A'_1$  & $\omega_{2}+\omega_{3}$ & $-0.89$ & 121 \\
$A'_{1}  \times A'_{1}$  & $A'_1$  & $2\omega_{3}$           & $-0.65$ & 145 \\
$A''_{2} \times A'_{1}$  & $A''_2$ & $\omega_{1}+\omega_{2}$ & $+0.60$ & 101 \\
$A''_{2} \times A'_{1}$  & $A''_2$ & $\omega_{1}+\omega_{3}$ & $-0.01$ & 113 \\ 
\noalign{\smallskip}
\hline
\end{tabular}
\end{table}

In addition to the six double bands of Fig.~\ref{vibvib}, other double bands appear
to be observed with $K^{P}=1^{-}$, in particular the four combination bands $%
E^{\prime }(8.84)\times A_{1}^{\prime }(6.72)$, $E^{\prime }(8.84)\times
A_{1}^{\prime }(7.19)$, $E^{\prime }(8.71)\times A_{1}^{\prime }(6.72)$, $%
E^{\prime }(8.71)\times A_{1}^{\prime }(7.19)$, observed at $%
K^{)}=1^{-}(11.98)$, $K^{P}=1^{-}(12.84)$, $K^{P}=1^{-}(13.12)$, $%
K^{P}=1^{-}(13.46)$. They are not reported here due to the uncertainty of
the even members, $2^{-}$, $4^{-}$, $\ldots$, of these bands.

\section{Other cluster configurations}

\begin{figure}
\centering
\setlength{\unitlength}{1pt}
\begin{picture}(320,435)(0,0)
\thicklines
\put(105,295) {\circle*{8}} 
\put( 60,310) {\circle*{8}}
\put(120,310) {\circle*{8}}
\put( 90,370) {\circle*{8}}
\put( 90,250) {\circle*{8}}
\put(140,260) {a)}
\multiput( 60,310)(4,0){15}{\circle*{1}}
\put( 60,310) {\line( 1, 2){30}}
\put( 60,310) {\line( 3,-1){45}}
\put( 60,310) {\line( 1,-2){30}}
\put(120,310) {\line(-1, 2){30}}
\put(120,310) {\line(-1,-2){30}}
\put(120,310) {\line(-1,-1){15}}
\put(105,295) {\line(-1, 5){15}}
\put(105,295) {\line(-1,-3){15}}
\multiput( 90,370)(0,-4){30}{\circle*{1}}
\put( 90,370) {\vector(0,1){40}}
\put( 95,395) {$z$-axis}
\put(210,310) {\circle*{8}} 
\put(240,310) {\circle*{8}}
\put(270,310) {\circle*{8}}
\put(240,370) {\circle*{8}}
\put(240,250) {\circle*{8}}
\put(300,260) {b)}
\put(240,250) {\line(-1, 2){30}}
\put(240,250) {\line( 1, 2){30}}
\put(240,250) {\line( 0, 1){120}}
\put(210,310) {\line( 1, 0){60}}
\put(240,370) {\line(-1,-2){30}}
\put(240,370) {\line( 1,-2){30}}
\put(225,301) {$a$}
\put(243,325) {$b$}
%
\put(200,200) {\circle*{8}} 
\put(220,200) {\circle*{8}}
\put(240,200) {\circle*{8}}
\put(260,200) {\circle*{8}}
\put(280,200) {\circle*{8}}
\put(200,200) {\line(1,0){80}}
\put(300,200) {c)}
\put(210, 50) {\circle*{8}} 
\put(260, 50) {\circle*{8}}
\put(190,100) {\circle*{8}}
\put(280,100) {\circle*{8}}
\put(235,136) {\circle*{8}}
\put(300, 70) {e)}
\put(210, 50) {\line( 1,0){50}}
\put(210, 50) {\line(-2,5){20}}
\put(260, 50) {\line( 2,5){20}}
\put(235,136) {\line(-5,-4){45}}
\put(235,136) {\line( 5,-4){45}}
\put( 60, 50) {\circle*{8}} 
\put(120, 70) {\circle*{8}}
\put( 80,170) {\circle*{8}}
\put(100,150) {\circle*{8}}
\put( 90,110) {\circle*{8}}
\put( 90, 60) {\circle*{4}}
\put( 90,160) {\circle*{4}}
\put(140,110) {d)}
\put( 90,160) {\vector(0,1){40}}
\put( 95,185) {$z$-axis}
\multiput( 60, 50)(5,0){8}{\circle*{1}}
\multiput( 60,150)(5,0){8}{\circle*{1}}
\multiput( 80,170)(5,0){8}{\circle*{1}}
\multiput( 60,150)(2,2){10}{\circle*{1}}
\multiput(100, 50)(2,2){10}{\circle*{1}}
\multiput(100,150)(2,2){10}{\circle*{1}}
\multiput(100, 50)(2,2){10}{\circle*{1}}
\put(100,150) {\line(-1, 1){20}}
\put( 60, 50) {\line( 3, 1){60}}
\put( 90,110) {\line(-1,-2){30}}
\put( 90,110) {\line( 3,-4){30}}
\put( 90,110) {\line(-1, 6){10}}
\put( 90,110) {\line( 1, 4){10}}
\multiput( 80, 70)(5,0){8}{\circle*{1}}
\multiput( 90, 60)(0,5){20}{\circle*{1}}
\multiput( 60, 50)(2,2){10}{\circle*{1}}
\multiput(100, 50)(-2,2){10}{\circle*{1}}
\multiput( 60,150)(3,1){20}{\circle*{1}}
\thinlines
\put( 50, 50) {\line(0,1){100}}
\put( 48, 50) {\line(1,0){4}}
\put( 48,150) {\line(1,0){4}}
\put( 60, 40) {\line(1,0){40}}
\put( 60, 38) {\line(0,1){4}}
\put(100, 38) {\line(0,1){4}}
\put(100, 40) {\line(1,1){20}}
\put(120, 58) {\line(0,1){4}}
\put( 40,100) {$h$}
\put( 80, 25) {$a$}
\put(110, 35) {$b$}
\end{picture}
\caption{Geometric configurations of five $\alpha$-particles:   
a) bi-pyramid (${\cal D}_{3h}$), 
b) planar diamond (${\cal D}_{2h}$ for $a \neq b$) 
and planar square (${\cal D}_{4h}$ for $a=b$), 
c) linear chain (${\cal D}_{\infty h}$), 
d) distorted tetrahedron (${\cal D}_{2d}$ for $a=b \neq h$) 
and tetrahedron (${\cal T}_d$ for $a=b=h$), 
and e) circular chain (${\cal D}_{5h}$).}
\label{geometries}
\end{figure}
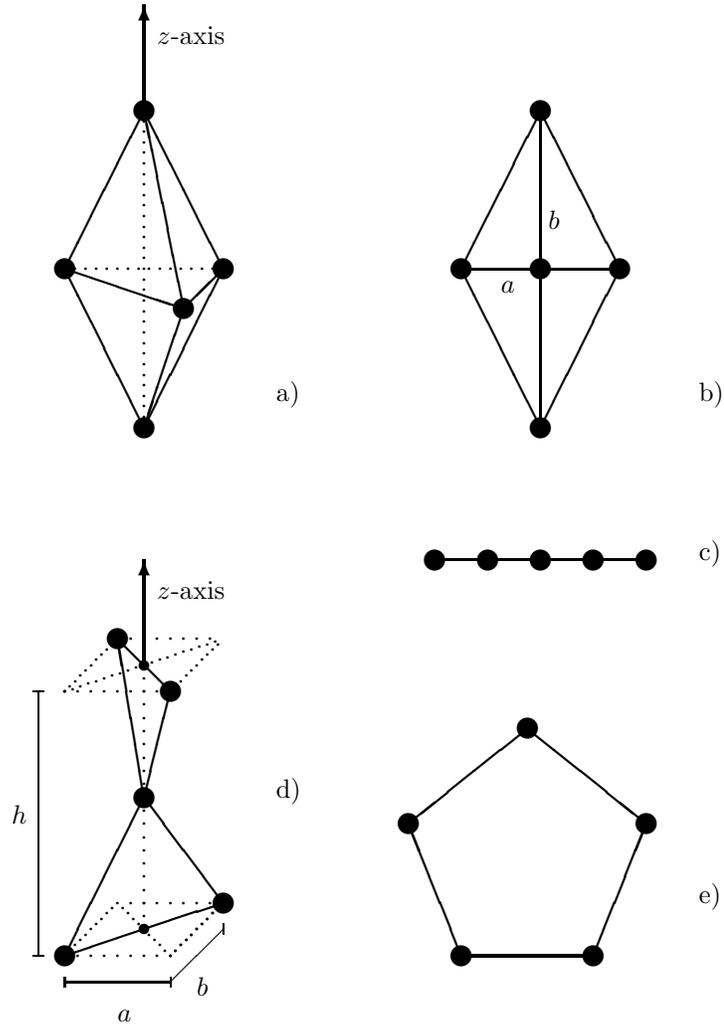

In addition to the bi-pyramidal configuration of Fig.~\ref{bipyramid}, 
$\alpha-^{12}$C(triangle)$-\alpha$, several others have been considered. In particular,
in 1971, Hauge, Willliams and Duffey \cite{hauge} suggested for $^{20}$Ne
the distorted body-centered tetrahedral configuration of Fig.~\ref{geometries}, $^{8}$Be$-\alpha-^{8}$Be, with symmetry $D_{2d}$. We have investigated in great
detail this configuration and derived all appropriate formulas which will be
given in a separate publication \cite{bijker-tobepublished}. However, this
configuration, apart from giving a good description of the $%
K^{P}=2^{-}(4.97)$ band as a hindered rotation, cannot describe
simultaneously the observed properties of the ground state band, and it does
not have the observed vibrations of the triangle $K^{P}=0^{+}(7.19)$, $%
K^{P}=1^{-}(8.84)$, $K^{P}=2^{+}(9.20)$. This configuration is expected at
much higher energy. A simple estimate can be done by observing that the $%
\alpha-^{12}$C$-\alpha$ configuration has 9 adjacent bonds, while the $^{8}$Be$-\alpha-^{8}$Be configuration has 6 adjacent bonds. The binding
energy of each configuration is proportional to the number of adjacent bonds 
\cite{hafstad}. In Fig.~\ref{geometries}, we also show another possible configuration, $\alpha-^{12}$C(linear)$-\alpha$, with symmetry $D_{2h}$. This configuration has 8 adjacent bonds and therefore is expected to be at a lower energy than the distorted body-centered tetrahedral structure. We have found no evidence for either configuration below an excitation energy of $12$ MeV.

Finally, Von Oertzen \cite{vonoertzen} suggested in 2001 a configuration $%
^{16}$O$-\alpha$. This configuration is similar to the bi-pyramidal
configuration, as one can see by replacing $^{16}$O by a tetrahedron \cite%
{bijker4}. The only difference is that the additional $\alpha $-particle may
be at a distance from the center greater than the $\alpha $-particle at the
vertex of the tetrahedron. Thus the $^{16}$O$-\alpha$ configuration has a $D_{3}$ symmetry rather than ${\cal D}_{3h}$, having lost the reflection symmetry on the horizontal $xy$-plane. Also, in \cite{vonoertzen} only the bands $K^{P}=0^{+}(0.0)$ and $K^{P}=0^{-}(5.79)$ were considered, while in the present paper all
vibrational bands are considered. A study of the $^{16}$O$-\alpha$ configuration 
was also done in Ref.~\cite{cseh} in terms of a nuclear vibron model.

\section{Quadrupole deformed structure}

A traditional collective description of $^{20}$Ne is in terms of a
quadrupole deformed ellipsoid with axial symmetry. Although this structure
describes the ground state band, $K^{P}=0^{+}(0.0)$, as accurately as the
bi-pyramid, it cannot account for the wealth of observed vibrational bands,
since it produces only two vibrations, a single degenerate vibration with $%
K^{P}=0^{+}$ ($\beta $-vibration) which can be associated with the band $%
K^{P}=0^{+}(6.72)$, and a doubly degenerate vibration with $K^{P}=2^{+}$ ($%
\gamma $-vibration) which can be associated with the band $K^{P}=2^{+}(9.20)$%
. Even if one adds to this description octupole degrees of freedom, one
cannot account for all observed vibrational states, in particular for the
vibrations of the triangle. It should be noted that the bi-pyramid can be
inscribed into an ellipsoid and thus some of the ellipsoidal features can be
obtained from those of the bi-pyramid. Specifically, the $\beta$-vibration is a combination of the two modes $\nu_2(A'_1)$ and $\nu_3(A'_1)$ of Fig.~\ref{modes}, and the $\gamma$-vibration is the $\nu_4(E')$ mode of Fig.~\ref{modes}.

\section{Microscopic description of $^{20}$Ne}

The early HF calculations of Brink \textit{et al.} \cite{brink2} suggested a
bi-pyramidal structure of $^{20}$Ne, in accordance with the findings of this
paper. Further calculations \cite{weiguny1,weiguny2} within the
Brink-Bloch model \cite{brink} have confirmed this finding. Modern DFT
calculations \cite{vretenar} also suggest, under appropriate conditions on
the nuclear forces, a bi-pyramidal structure. These calculations 
describe well the ground state properties of $^{20}$Ne, but they are unable
to calculate its excitation spectrum.

The shell model in its various forms is the appropriate model to describe
the microscopic structure of $^{20}$Ne. Many calculations have been done
throughout the years. Calculations in the $sd$-shell with large effective
charges, $e_{n}=0.5e$ and $e_{p}=1.5e$, describe well the ground state band.
Enlarging the space to the $pf$-shell with large octupole effective charges,
they describe well also some of the negative parity states. Introducing an
Elliott $SU(3)$ basis \cite{elliott}, many observed bands can be
classified in terms of representations of $SU(3)$ \cite{draayer1,anantaraman}. In particular, the ground state band $K^{P}=0^{+}(0.0)$ is
assigned to the representation $(\lambda,\mu)K=(8,0)0$ of
$SU(3)$, the low-lying band $K^{P}=2^{-}(4.97)$ is assigned to the
representation $(8,2)2$, the band $K^{P}=0^{-}(5.79)$ to the representation $%
(9,0)0$, and the band $K^{P}=0^{+}(6.72)$ to the representation $(4,2)0$ and
similarly for other bands. An enlargement to the $Sp(6,R)$ basis allows also
to eliminate the need for effective charges. It would be interesting to see
the extent to which the assignments of this paper can be accommodated within 
$SU(3)$ and $Sp(6,R)$. Finally, recently, an accurate parametrization of
$pfsd$ matrix elements has been developed \cite{caurier}. It would be of great
interest to see whether a modern large-scale shell-model calculation can
describe the wealth of available data in $^{20}$Ne.

\section{Summary and conclusions}

In this article, we have analyzed in great detail the available experimental
data in $^{20}$Ne and shown that they can, to a good approximation, be
decribed in terms of the bi-pyramidal cluster configuration suggested by
Brink in 1970 \cite{brink2}. Ground state properties can be described in
terms of only two geometric parameters $\beta _{1}$ and $\beta _{2}$ with an
accuracy comparable to that of microscopic calculations. Surprisingly, all
observed $K^{P}$ bands with band-heads up to an excitation energy of about $%
12$ MeV can be accounted for in terms of the vibrationless ground state
band, of the nine vibrational modes expected on the basis of ${\cal D}_{3h}$
symmetry (3 singly degenerate and 3 doubly degenerate), and of six of the
double vibrational bands expected on the basis of ${\cal D}_{3h}$ symmetry. The
only band with anomalous behavior out of the 16 analyzed is the band with $%
K^{P}=2^{-}(4.79)$ which occurs at a much lower energy than expected ($\sim
9 $ MeV).

We believe that the study presented here provides strong evidence for a
quasi-molecular structure of $^{20}$Ne with ${\cal D}_{3h}$ symmetry. $^{20}$Ne
appears to be another example of the simplicity in complexity program, in
which simple spectroscopic features arise out of a complex many-body system.
It would be of great interest to see whether these simple features can also
be obtained in microscopic calculations, such as large-scale shell-model
calculations.

\section{Acknowledgements}

This work was supported in part by research grant IN101320 from PAPIIT-DGAPA.

\end{document}